\documentclass{aa}
\usepackage{graphicx}

\begin{document}

\title{Order and chaos in the local disc stellar kinematics induced by the
Galactic bar}

\author{R. Fux}

%\offprints{R. Fux} (only if more than one author)

\institute{Research School of Astronomy and Astrophysics, Australian National
           University, Mount Stromlo Observatory, Cotter Road,
           Weston Creek ACT 2611, Australia\\
           \email{fux@mso.anu.edu.au}}

\date{Received xx / Accepted xx}

%%%%%%%%%%%%%%%%%%%%%%%%%%%%%%%%%%%%%%%%%%%%%%%%%%%%%%%%%%%%%%%%%%%%%%
\abstract{
The Galactic bar causes a characteristic splitting of the disc phase space
into regular and chaotic orbit regions which is shown to play an important
role in shaping the stellar velocity distribution in the Solar neighbourhood.
A detailed orbital analysis within an analytical 2D rotating barred potential
reveals that this splitting is mainly dictated by the value of the Hamiltonian
$H$ and the bar induced resonances. In the $u-v$ velocity plane at fixed space
position, the contours of constant~$H$ are circles centred on the local solid
rotation velocity of the bar frame and of radius increasing with $H$. For
reasonable bar strengths, the contour $H=H_{12}$ corresponding to the
effective potential at the Lagrangian points $L_{1/2}$ marks the average
transition from regular to chaotic motion, with the majority of orbits being
chaotic at $H>H_{12}$. On top of this, the resonances generate an alternation
of regular and chaotic orbit arcs opened towards lower angular momentum and
asymmetric in $u$ for space positions away from the principal axes of the bar.
Test particle simulations of exponential discs in the same potential and a
more realistic high-resolution 3D \mbox{$N$-body} simulation reveal how the
decoupled evolution of the distribution function in the two kind of regions
and the process of chaotic mixing lead to overdensities in the $H\ga H_{12}$
chaotic part of the disc velocity distributions outside corotation.
In particular, for realistic space positions of the Sun near or slightly
beyond the outer Lindblad resonance and if $u$ is defined positive towards
the anti-centre, the eccentric quasi-periodic orbits trapped around the stable
$x_1(1)$ orbits -- i.e. the bar-aligned closed orbits which asymptotically
become circular at larger distances -- produce a broad $u\la 0$ regular arc in
velocity space extending within the $H>H_{12}$ zone, whereas the corresponding
$u\ga 0$ region appears as an overdensity of chaotic orbits forced to avoid
that arc. This chaotic overdensity
provides an original interpretation, distinct from the anti-bar elongated
quasi-periodic orbit interpretation proposed by Dehnen (\cite{D5}), for the
prominent stream of high asymmetric drift and predominantly outward moving
stars clearly emerging from the Hipparcos data. However, the most appropriate
interpretation for this stream remains uncertain.
The effects of spiral arms and of molecular clouds are also briefly discussed
within this context.
%%%%%%%%%%%%%%%%%%%%%%%%%%%%%%%%%%%%%%%%%%%%%%%%%%%%%%%%%%%%%%%%%%%%%%
}
\maketitle
%%%%%%%%%%%%%%%%%%%%%%%%%%%%%%%%%%%%%%%%%%%%%%%%%%%%%%%%%%%%%%%%%%%%%%

\section{Introduction}
%%%%%%%%%%%%%%%%%%%%%%
%
\par The kinematics of disc stars in the Solar neighbourhood displays several
long known properties, such as the increase of velocity dispersion with age,
the tendency of young stars to appear in moving groups or streams, and the
classical vertex deviation affecting stars with asymmetric drift down to 
$\sim 25$~km\,s$^{-1}$ relative to the Sun and mainly owing to the Hyades and
Sirius streams. Disc heating is traditionally attributed to the diffusion of
stars by transient spiral arms or by massive compact objects like molecular
clouds, the streams to dissolving ensembles of stars born at the same place,
and the vertex deviation to local gravitational perturbations like spiral
arms or local departures from a steady state.
\par
\begin{figure*}
\includegraphics[width=17.5cm]{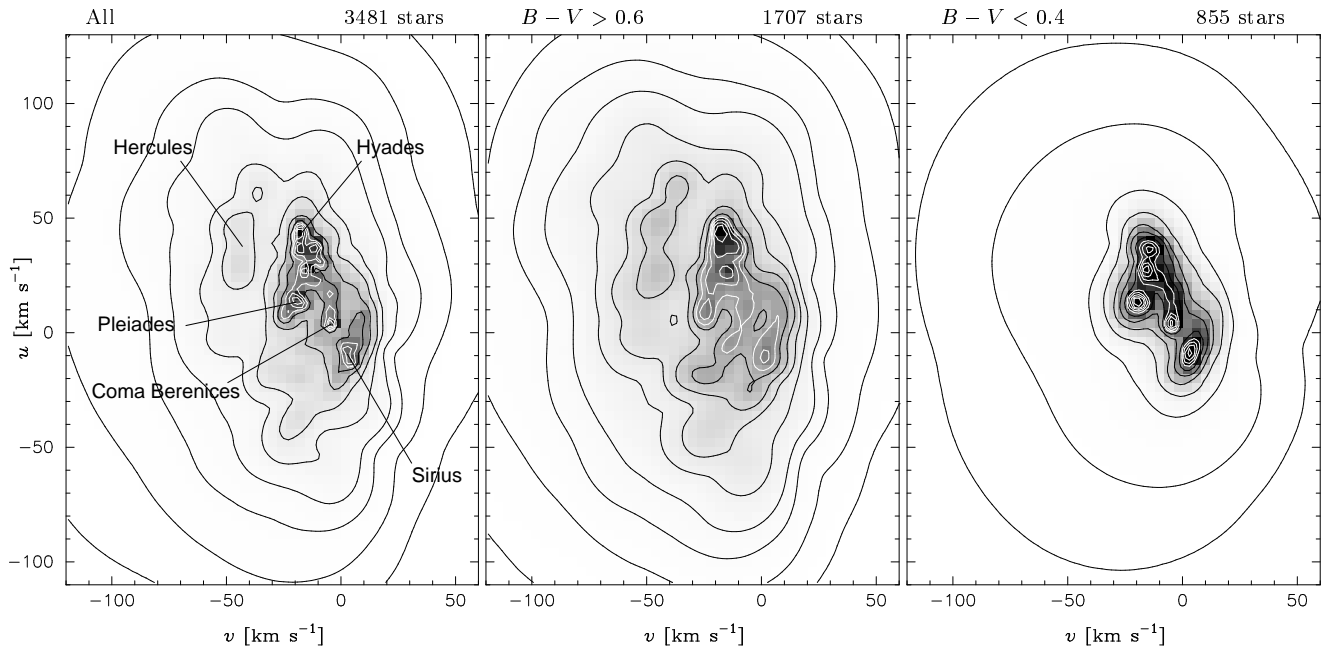}
\caption{\small Heliocentric velocity distribution in the $u-v$ plane of all the
Hipparcos single stars with $\sigma(\pi)/\pi<0.1$, $d<100$~pc and radial
velocities in the Hipparcos Input Catalogue (left) and of the sub-samples with
$B-V>0.6$ (middle) and $B-V<0.4$ (right). For the sake of comparison, the
contours are as in Dehnen (\cite{D1}), containing $2$, $6$, $12$, $21$, $33$,
$50$, $68$, $80$, $90$, $95$, $99$ and $99.9$ percent of all stars. The
diagram for the full sample is exactly the same as in Fux (\cite{F3}), except
for a different labelling of the contours.}
\label{obs}
\end{figure*}

 Beside these properties, the local disc velocity distribution also
betrays a broad stream of low angular momentum and mainly outward moving stars
with a mean heliocentric asymmetric drift $s\approx 45$~km\,s$^{-1}$, i.e.
typical of the thick disk (Gilmore et al. \cite{GWK}), which hereafter will be
referred to as the ``Hercules'' stream, according to the comoving Eggen group
$\zeta$ Herculis (Skuljan et al. \cite{SHC}). The mean outward motion of stars
with high asymmetric drift, also known as the ``$u$-anomaly'' and seen up to
over $s=100$~km\,s$^{-1}$ in metal rich samples (Raboud et al. \cite{RGMFS})
and in Mira variables with period between $145$ and $200$ days (Feast \&
Whitelock \cite{FW}), is already apparent in early stellar kinematical samples
(Eggen \cite{E}; Woolley et al. \cite{WEPP}) and was recognised long ago by
Mayor (\cite{M}), but the clearest evidence for the Hercules stream comes from
the Hipparcos proper motions combined with (Fig.~\ref{obs}) or without (Dehnen
\cite{D1}) available radial velocities. This stream is very likely to have a
dynamical origin because its stars are older than $\sim 2$ Gyr (Caloi et al.
\cite{CCD}) and present a wide range of metallicities (Raboud et al.
\cite{RGMFS}).
\par The existence of the Hercules stream is most probably related to the
influence of the Galactic bar. It is now indeed widely accepted that the
Milky~Way is a barred galaxy, as are the majority of disc galaxies. Evidence
for the bar comes from longitudinal asymmetry in the bulge surface photometry
(e.g. Blitz \& Spergel \cite{BS}; Binney et al. \cite{BGS}), star counts (e.g.
Nakada et al. \cite{NDH}; Nikolaev \& Weinberg \cite{NW}; Stanek et al.
\cite{SU}), interpretation of the observed gas kinematics in the central few
kpc (Binney et al. \cite{BGSBU}; Englmaier \& Gerhard \cite{EG}; Fux
\cite{F2}; Weiner \& Sellwood \cite{WS}), large microlensing optical depths
towards the Galactic bulge (Paczynski et al. \cite{PSU}; Kuijken \cite{KK};
Gyuk \cite{GY}; Alcock et al. \cite{AAA}) and possibly inner stellar
kinematics (Sevenster et al. \cite{SSVF}; see also Gerhard \cite{OG} for a
recent review). Although still not very well constrained, the most quoted
values for the main bar parameters are an in-plane inclination angle with
respect to the Galactic centre direction
$\varphi\approx 15^{\circ}-45^{\circ}$, with the near side of the bar in the
first Galactic quadrant, and a corotation radius
$R_{\hbox{\tiny CR}}\approx 3.5-5$~kpc.
\par Barred \mbox{$N$-body} models of the Milky Way produce a mean outward
motion of disc particles at realistic positions of the Sun relative to the bar
(Fux et al. \cite{FMP}; Raboud et al. \cite{RGMFS}), but the precise bar
induced dynamical process leading to the observed kinematical properties of
the Hercules stream is still a matter of debate. Dehnen (\cite{D4}, \cite{D5}
-- hereafter D2000) relates this stream and the main mode of high angular
momentum stars in the observed velocity distribution to the coexistence near
the outer Lindblad resonance (OLR) of two distinct types of periodic orbits
replacing the circular orbit close to the OLR in a rotating barred potential,
i.e. the same idea introduced by Kalnajs (\cite{K}) to explain the Hyades and
Sirius streams. Linear theory indeed predicts that the orientation of orbits
closing in the bar rotating frame changes across the main resonances
associated with the bar (Binney \& Tremaine \cite{BT}). In particular,
periodic orbits outside and inside the OLR radius are respectively elongated
along the major and minor axis of the bar, and both types of orbits, as well
as the quasi-periodic orbits trapped around these orbits, can overlap in space
near the OLR. According to D2000, the Hercules stream and the main velocity
mode, respectively ``OLR'' and ``LSR'' mode in his terminology, result from
the anti-bar and bar elongated orbits respectively, and the valley between the
two modes from off-scattered stars on unstable OLR orbits. Raboud et al.
(\cite{RGMFS}), on the other hand, suggest that the Hercules stream involves
stars merely on chaotic orbits and susceptible to cross the corotation radius
and wander throughout the Galaxy, but do not explicitly justify why such stars
should move outwards on the average in the Solar neighbourhood. One motivation
for this interpretation is that of order $10\%$ of the particles in
\mbox{$N$-body} models of barred galaxies indeed follow such orbits (e.g.
Pfenniger \& Friedli \cite{PF}).
\par This paper investigates how the barred potential of the Milky Way divides
the phase space of the stellar disc into regions of regular and chaotic motion
and how this segregation may explain some properties of the observed local
stellar kinematics and in particular help to clarify the real nature of the
Hercules stream. The investigation is first performed in details using the
same analytical two-dimensional rotating barred potential as in D2000 and then
complemented with the results from a more \mbox{realistic} high-resolution
three-dimensional \mbox{$N$-body} simulation.
\par The structure of the paper is as follow: Section~\ref{hip} briefly
presents the observed stellar velocity distribution in the Solar neighbourhood
and some further informations about the Hercules stream. Section~\ref{hampot}
recalls a dynamical classification of orbits in rotating barred potentials
based on the Jacobi integral and determines the location in local velocity
space of the class of orbits that may cross the corotation radius.
Section~\ref{wpot} describes the analytical barred potential adopted in the 2D
study and Sect.~\ref{pero} the main periodic orbits supported by this
potential outside corotation. Section~\ref{liaexp} derives the associated
regular and chaotic regions in velocity space as a function of space position
relative to the bar. Section~\ref{testpart} presents the velocity
distributions at the same space positions resulting from test particle
simulations and examines the role of chaos in shaping these distributions.
Section~\ref{incond} shows how the derived velocity distributions depend on
the initial conditions of the simulations and Sect.~\ref{restock} how the
particles initially on OLR orbits eventually contribute to these distributions.
Section~\ref{nbody} gives the results inferred from the 3D \mbox{$N$-body}
simulations. Section~\ref{compar} makes a quantitative comparison of the model
velocity distributions with the observed one and discusses the most likely
origin of the Hercules stream. Finally, Sect.~\ref{concl} sums up.

\section{Observed local velocity distribution}
%%%%%%%%%%%%%%%%%%%%%%%%%%%%%%%%%%%%%%%%%%%%%%
\label{hip}

There are several attempts to recover the velocity distribution of stars in
the Solar neighbourhood from the Hipparcos data published in the recent
literature (e.g. Dehnen \cite{D1} and Skuljan et al. \cite{SHC} for all
stellar types; Chereul et al. \cite{CCB} and Asiain et al. \cite{AFTC} for
early-type stars). The main features of the $u-v$ distribution are illustrated
in Figure~\ref{obs} and the mean velocities of the highlighted streams are
listed in Table~\ref{stream}. Throughout this paper, $v$ and $u$ respectively
stand for the azimuthal and radial velocity components, with positive values
towards galactic rotation and towards the Galactic anti-centre, and $w$ for
the vertical velocity component.
\par The main sample selected for this figure is built from the 3481 single
stars of the Hipparcos Catalogue with relative errors on parallaxes less than
$10$\%, distances less than $100$~pc, and given radial velocities in the
Hipparcos Input Catalogue. Here, an entry of the Hipparcos Catalogue is
considered as a single star if the CCDM identifier and Multiple System Annex
flag (fields H55 and H59 respectively) are void, the number of components
(field H58) is $1$ and the solution quality (field~H61) is different from 'S'.
Two disjoint sub-samples are isolated from this main sample, the first one
restricted to the 1707 stars with $B-V>0.6$, representing stars which are
older on the average than the stars in the full sample, and the second one
to the 855 stars with $B-V<0.4$, representing essentially main sequence stars
which are younger than 2~Gyr. The diagrams are derived using the adaptative
kernel method described in Skuljan et al. (\cite{SHC}), with an average
smoothing length $h=16$~km\,s$^{-1}$ for the $B-V>0.6$ sub-sample, and
$h=10$~km\,s$^{-1}$ for the other samples.
\begin{table}
\centering
\caption{\small Mean heliocentric velocities of some stellar streams in the Solar
neighbourhood. The velocity components are estimated from the left frame in
Fig.~\ref{obs}, except those for the Arcturus stream which refer to figure~3
of Dehnen (\cite{D1}).}
\begin{tabular}{lrr} \hline \vspace*{-.35cm} \\
Stream & $v\;[\,$km\,s$^{-1}]$ & $u\;[\,$km\,s$^{-1}]$\\ \hline
\vspace*{-.35cm} \\
Coma Berenices &   $-4$~~~~~~~ &  $3$~~~~~~~ \\
Sirius         &    $3$~~~~~~~ & $-9$~~~~~~~ \\
Hyades         &  $-18$~~~~~~~ & $42$~~~~~~~ \\
Pleiades       &  $-19$~~~~~~~ & $13$~~~~~~~ \\
Hercules       &  $-45$~~~~~~~ & $35$~~~~~~~\vspace*{.1cm}\\
Arcturus       & $-110$~~~~~~~ & $16$~~~~~~~ \\ \hline
\label{stream}
\end{tabular}
\end{table}
\par The reader should be warned that stellar samples built this way are
kinematically biased in the sense that radial velocities are predominantly
known for high proper-motion stars (Binney et al.~\cite{BDHMP}; Skuljan et
al.~\cite{SHC}). Moreover, the completeness of the Hipparcos Catalogue depends
on Galactic latitude, so that the effects of such a bias are even further
complicated by the anisotropic local velocity distribution. Nevertheless, the
resulting velocity distributions closely resemble the asserted unbiased
distributions derived by Dehnen (\cite{D1}), suggesting that kinematical
biases do not severely affect our diagrams.
\par Figure~\ref{obs} nicely confirms that the Hercules stream involves merely
old disc stars. According to D2000, roughly 15\% of the Hipparcos stars with
$B-V>0.6$ belong to this stream, but this is likely an underestimate of the
corresponding fraction among local old disc stars because such a colour range
is still contaminated by young stars which contribute negligibly to the stream
and because the Hipparcos catalogue is biased towards young stars. The average
luminosity of stars in this catalogue indeed increases with distance and the
catalogue essentially covers the vertical region of the Galactic plane where
the fraction of young stars is largest.

\section{Effective potential and Jacobi integral}
%%%%%%%%%%%%%%%%%%%%%%%%%%%%%%%%%%%%%%%%%%%%%%%%%%
\label{hampot}
In a rigid potential $\Phi(\vec{x})$ rotating at a constant frequency
$\Omega_{\rm P}$ about the \mbox{$z$-axis}, the Hamiltonian of a test particle
expressed in the rotating frame writes:
\begin{equation}
H(\vec{x},\dot{\vec{x}})=\frac{1}{2}\dot{\vec{x}}^2+\Phi_{\rm eff}(\vec{x}),
\label{ham}
\end{equation}
where $\Phi_{\rm eff}(\vec{x})=\Phi(\vec{x})-\frac{1}{2}\Omega_{\rm P}^2
(x^2+y^2)$ is the effective potential. If $\Phi(\vec{x})$ is non-axisymmetric
and $\Omega_{\rm P}\neq 0$, the energy $E$ and the \mbox{$z$-component} of the
angular momentum $L_z$ are not conserved individually, and the only known
classical integral of motion generally is the value of the Hamiltonian
$H=E-\Omega_{\rm P}L_z$, known as the Jacobi integral. Since $\dot{\vec{x}}^2$
must be positive, this integral restricts the motion of a particle to the
space region where $\Phi_{\rm eff}(\vec{x})<H$.
\begin{figure}
\includegraphics[width=8cm]{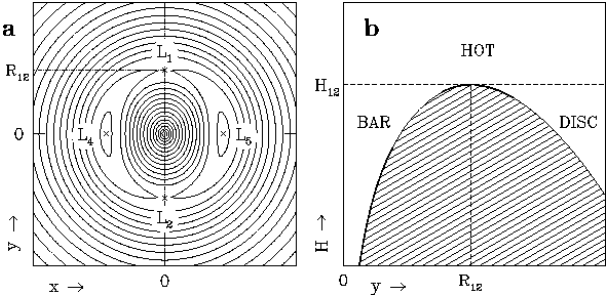}
\caption{\small {\bf a)} Effective potential in the $x-y$ plane of a barred disc
model corresponding to Eq.~(\ref{pot}) with $F=0.10$. The bar is along the
\mbox{$y$-axis} and the spacing between the contours increases by a factor
$1.2$ towards lower $\Phi_{\rm eff}$. The crosses indicate the Lagrangian
points $L_{1/2}$ (on the \mbox{$y$-axis}) and $L_{4/5}$ (on the
\mbox{$x$-axis}). {\bf b)} Effective potential along the \mbox{$y$-axis}
(thick line) and the three main classes of orbits related to the conservation
of the Jacobi integral. The shaded area below the curve is forbidden for
orbits with $H<H_{12}$.}
\label{poteff}
\end{figure}
\par In a rotating barred potential, the contours of effective potential in
the plane of symmetry $z=0$ look like a volcano with a sinusoidal crest, the
extrema of which defining the locations of the Lagrangian points $L_{1/2}$ and
$L_{4/5}$, corresponding respectively to the saddle points and maxima of
$\Phi_{\rm eff}$ on the major and minor axis of the bar (Fig.~\ref{poteff}a).
Two critical values of the Hamiltonian are associated with stars corotating at
these points, namely $H_{12}\equiv \Phi_{\rm eff}(L_{1/2})$ and
$H_{45}\equiv \Phi_{\rm eff}(L_{4/5})$. The first of them can be used to
classify stellar orbits into three dynamical categories (Sparke \& Sellwood
\cite{SS}; Pfenniger \& Friedli \cite{PF}): the {\it bar orbits} and {\it disc
orbits} with $H<H_{12}$, which cannot cross the $H_{12}$ contour and are
therefore confined inside and outside corotation respectively, and the {\it
hot orbits} with $H\geq H_{12}$, which are susceptible to cross the corotation
barrier and explore all space except a small region around $L_{4/5}$ if
$H<H_{45}$ (Fig.~\ref{poteff}b). Stars with $H_{12}<H<H_{45}$ cannot cross the
corotation radius at all azimuth and may therefore more likely be locked
during several orbital periods on either side of corotation.
\par
\begin{figure}
\includegraphics[width=8cm]{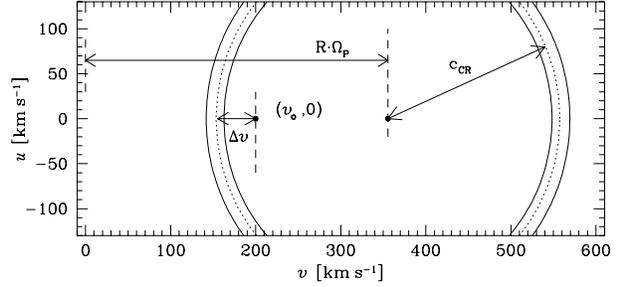}
\caption{\small Contours of constant Hamiltonian in the $u-v$ plane of a realistic 2D
barred model (Eq.~(\ref{pot}) with $R_{\hbox{\tiny CR}}/R_{\circ}=4.5/8$,
$v_{\circ}=200$~km\,s$^{-1}$ and $F=0.20$). The velocities are relative to an
inertial frame. The inner and outer solid circles give the $H_{12}$ and
$H_{45}$ contours respectively, and the dotted circle is the axisymmetric
limit of these contours.}
\label{uvham}
\end{figure}

 In the Solar neighbourhood, located confidently beyond corotation, only
stars from the disc and hot populations are observed. Since these stars share
about the same $\Phi_{\rm eff}$ if not too far from the Galactic plane, their
$H$-values depend mainly on the velocities and thus one expects that the two
populations occupy different regions in local velocity space. If $v$, $u$ and
$w$ are measured with respect to the Galactic centre, Eq.~(\ref{ham})
transforms into:
\begin{equation}
(v-R_{\circ}\Omega_{\rm P})^2+u^2+w^2=2(H-\Phi_{\rm eff}^{\circ}),
\label{cont}
\end{equation}
where $R_{\circ}$ is the galactocentric distance of the Sun and
$\Phi_{\rm eff}^{\circ}$ the local effective potential. Thus the contours of
constant Hamiltonian in velocity space are spheres centred on
$(v,u,w)=(R_{\circ}\Omega_{\rm P},0,0)$ and of radius
$\sqrt{2(H-\Phi_{\rm eff}^{\circ})}$ increasing with $H$. Stars on disc and
hot orbits are respectively those inside and outside the $H_{12}$ sphere. If
the vertical dimension is neglected, these spheres become circles with the
same properties. In the axisymmetric limit and for a flat rotation curve of
circular velocity $v_{\circ}$, the radius $c_{\hbox{\tiny CR}}$ of the
$H_{12}=H_{45}$ contour and the low azimuthal velocity $\Delta v$ relative to
$v_{\circ}$ at which this contour crosses the $u=0$ axis are then given by:
\begin{eqnarray}
c_{\hbox{\tiny CR}}^2 & \equiv & 2[H_{12}-\Phi_{\rm eff}(R_{\circ})] \nonumber
\label{rad} \\
& = & v_{\circ}^2\left[\ln{\left(\frac{R_{\hbox{\tiny CR}}}{R_{\circ}}
\right)^2}+\left(\frac{R_{\circ}}{R_{\hbox{\tiny CR}}}\right)^2-1\right], \\
\Delta v & = & v_{\circ}\left(\frac{R_{\circ}}{R_{\hbox{\tiny CR}}}-1\right)
-c_{\hbox{\tiny CR}},
\label{dV}
\end{eqnarray}
where $R_{\hbox{\tiny CR}}=v_{\circ}/\Omega_{\rm P}$ is the corotation radius
(see Fig.~\ref{uvham}). For $R_{\hbox{\tiny CR}}/R_{\circ}=4.5/8$ and
$v_{\circ}=200$~km\,s$^{-1}$, one gets
$-\Delta v=0.227v_{\circ}\approx 45$~km\,s$^{-1}$, which coincides with the
mean heliocentric asymmetric drift of the Hercules stream\footnote{It is
implicitly assumed here that the azimuthal velocity of the Sun is close to the
circular velocity of the axisymmetric part of the Galactic potential. This is
probably correct within $5-10$~km\,s$^{-1}$, as will be argued in
Sect.~\ref{compar}.}. Note however that this simple approximation is not truly
a lower limit to the asymmetric drift of stars on hot orbits for several
reasons: a non-zero $w$ velocity component defines two circles on the $H_{12}$
sphere with a reduced projected radius on the $u-v$ plane (small effect, of
order $1$~km\,s$^{-1}$ for $|w|=20$~km\,s$^{-1}$), the presence of a bar
lowers the effective potential at $L_{1/2}$ (larger effect, of order
$5-10$~km\,s$^{-1}$), and finally $|\Delta v|$ is smaller if $u\neq 0$. Hence
most stars in the Hercules stream are likely to fall outside the $H_{12}$
sphere and therefore may belong to the hot population.
\par From Eqs.~(\ref{rad}) and~(\ref{dV}), it also follows that the radius
$c_{\hbox{\tiny CR}}$ and the velocity separation $|\Delta v|$ increase for
larger galactocentric distances relative to $R_{\hbox{\tiny CR}}$. In
particular, whatever the strength of the bar, one can always increase the
fraction of the Hercules stream falling in the hot orbit region by reducing
the value of $R_{\circ}/R_{\hbox{\tiny CR}}$.

\section{Working potential}
%%%%%%%%%%%%%%%%%%%%%%%%%%%
\label{wpot}

The analytical 2D barred potential adopted for the orbital structure analysis
and the test particle simulations is the same as in D2000:
\begin{equation}
\Phi(R,\phi)=\Phi_{\circ}(R)+\Phi_{\rm b}(R,\phi),
\label{pot}
\end{equation}
with
\vspace*{-0.05cm}
\begin{eqnarray}
\hspace*{1cm}
\Phi_{\circ}(R) & = & v_{\circ}^2\ln{R}, \label{axi} \\
\Phi_{\rm b}(R,\phi) & = & \frac{1}{2}Fv_{\circ}^2\cos{(2\phi)}\left\{
\begin{array}{cl}
2-\left(\frac{R}{a}\right)^3  & R\leq a\vspace*{.1cm}\\
\left(\frac{R}{a}\right)^{-3} & R\geq a.
\end{array}\right.
\end{eqnarray}
This represents the sum of an axisymmetric potential $\Phi_{\circ}$ with
constant circular velocity $v_{\circ}$ and a barred potential $\Phi_{\rm b}$
falling off as a quadrupole at $R\geq a$. The inner and outer parts of the
latter component are described by two distinct functions which connect
together at $R=a$ such as to ensure continuous potential and forces. The bar
major axis is taken to coincide with the \mbox{$y$-axis}, contrary to the
convention in D2000. The parameter $F$ is the bar strength, defined as the
maximum azimuthal force on the circle of radius $a$ divided by the radial
force of the axisymmetric part of the potential at the same radius (in
absolute value). It is related to Dehnen's parameter $\alpha$ by
$F=8.89\alpha$. The potential is rotating at a constant pattern speed
$\Omega_{\rm P}$ such as to place corotation at $R_{\hbox{\tiny CR}}=1.25 a$,
in agreement with numerical simulations and analyses of observations in
early-type barred galaxies if $a$ is associated with the bar semi-major axis
(e.g. Elmegreen \cite{BE}). Unlike D2000, only flat rotation curve models will
be examined. In this case, the OLR and corotation radii are related via
$R_{\hbox{\tiny OLR}}=(1+1/\sqrt{2})R_{\hbox{\tiny CR}}$, and a value of
$R_{\circ}/R_{\hbox{\tiny OLR}}=1.1$ corresponds to a corotation radius
$R_{\hbox{\tiny CR}}=4.26$~kpc if $R_{\circ}=8$~kpc. Some considerations in
the case of a non-constant rotation curve can be found in Sect.~\ref{compar}.
\par The next sections present a study of the periodic orbits outside
corotation in the adopted rotating barred potential, identify the regular and
chaotic regions in phase space associated with this potential, and discuss how
stars may populate the available orbits. All orbits in these sections are
integrated in double precision using an 8 order Runge-Kutta-Fehlberg algorithm
(Fehlberg \cite{RKF}). Two values of the bar strength will be considered,
$F=0.10$ and $F=0.20$. The larger value corresponds to a rather strong bar
(see Sect.~\ref{nbody} for a quantification with respect to real galaxies),
but has the advantage to clearly point out the effect of chaos in the test
particle simulations. Some of the key results for the intermediate case
$F=0.15$ are presented in Fux (\cite{F4}). For comparison, Dehnen's
simulations were done in the range $F=0.062-0.116$.

\section{Periodic orbits}
%%%%%%%%%%%%%%%%%%%%%%%%%
\label{pero}

Much of the orbital structure in a system can be assessed from the study of
its periodic orbits, which sometimes are considered as the skeleton of the
system. While periodic orbits have been widely investigated in 2D and 3D
within bars, only few papers (Athanassoula et al. \cite{ABMP}; Contopoulos \&
Grosbol \cite{CG}; Sellwood \& Wilkinson \cite{SW}) discuss them in discs
surrounding bars.
\begin{figure}
\includegraphics[width=8cm]{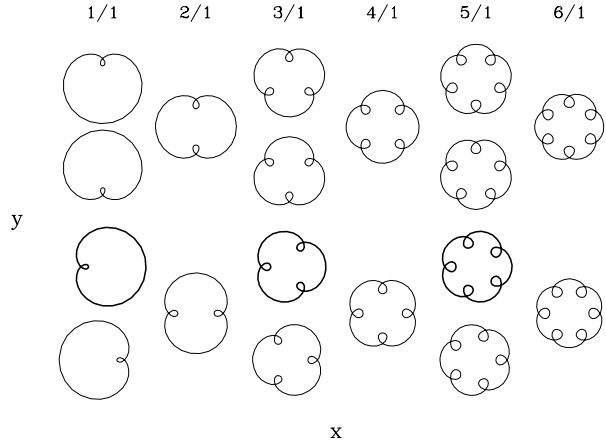}
\caption{\small Distinct looped periodic orbits in the axisymmetric
potential~$\Phi_{\circ}$ of Eq.~(\ref{axi}) which close after one rotation in
the rotating frame and are symmetric with respect to the $x$- and/or
\mbox{$y$-axes}. The orbits all have
$(H-H_{\hbox{\tiny OLR}})/v_{\circ}^2=0.244$ and are not drawn at the same
relative scale. For cross-identification, the line thickness of the orbits is
the same as for the portion of the characteristic curves in Fig.~\ref{char}a
to which they belong. Orbits in the lower and upper halfs of the diagram give
rise to respectively stable and unstable orbits at low eccentricities when the
bar is added.}
\label{orb0}
\end{figure}
\par A good approach to investigate periodic orbits in a 2D rotating barred
potential is to start with the axisymmetric limit. In this case, the only
orbits which close whatever the value of $\Omega_{\rm P}$ are the circular
orbits. All other bound orbits can be thought as a libration motion around
these orbits and look like a rosette which never closes, except for some
exceptional potentials like a point mass with $\Omega_{\rm P}=0$, or at
resonances, where the radial and azimuthal frequencies $\omega_R$ and
$\omega_{\phi}$ satisfy the relation:
\begin{equation}
n_{\phi}\omega_R=n_R(\omega_{\phi}-\Omega_{\rm P}),
\end{equation}
with integer values of $n_R\geq 0$ and $n_{\phi}$. In the rotating frame, an
$n_R/n_{\phi}$ resonant orbit closes after $n_R$ radial oscillations and
$|n_{\phi}|$ orbital periods. Outside corotation, $n_{\phi}$ is negative and
one may speak of outer $n_R/|n_{\phi}|$ resonances. This paper discusses only
outer resonances and the minus sign in their labelling will be omitted. While
in the axisymmetric case the orientation of the resonant orbits is arbitrary,
the virtual introduction of a bar will retain only those orbits which are
reflection symmetric with respect to (at least one of) the bar principal axes,
say the $x$- and \mbox{$y$-axes}. Figure~\ref{orb0} displays several orbits of
this kind with $|n_{\phi}|=1$. There exists an infinity of resonant orbits
which accumulate at corotation as $n_R\rightarrow \infty$. Those with $n_R>6$
or closing after more than one rotation will not be considered in this paper.
For even $n_R$, there exists two distinct resonant orbits for each value of
the Hamiltonian, depending on whether the \mbox{$x$-axis} coincides with
orbital apocentre or pericentre. For odd $n_R$, there are twice as many
solutions because the orbits are no longer reflection symmetric with respect
to both axes.
\begin{figure*}
\includegraphics[width=17cm]{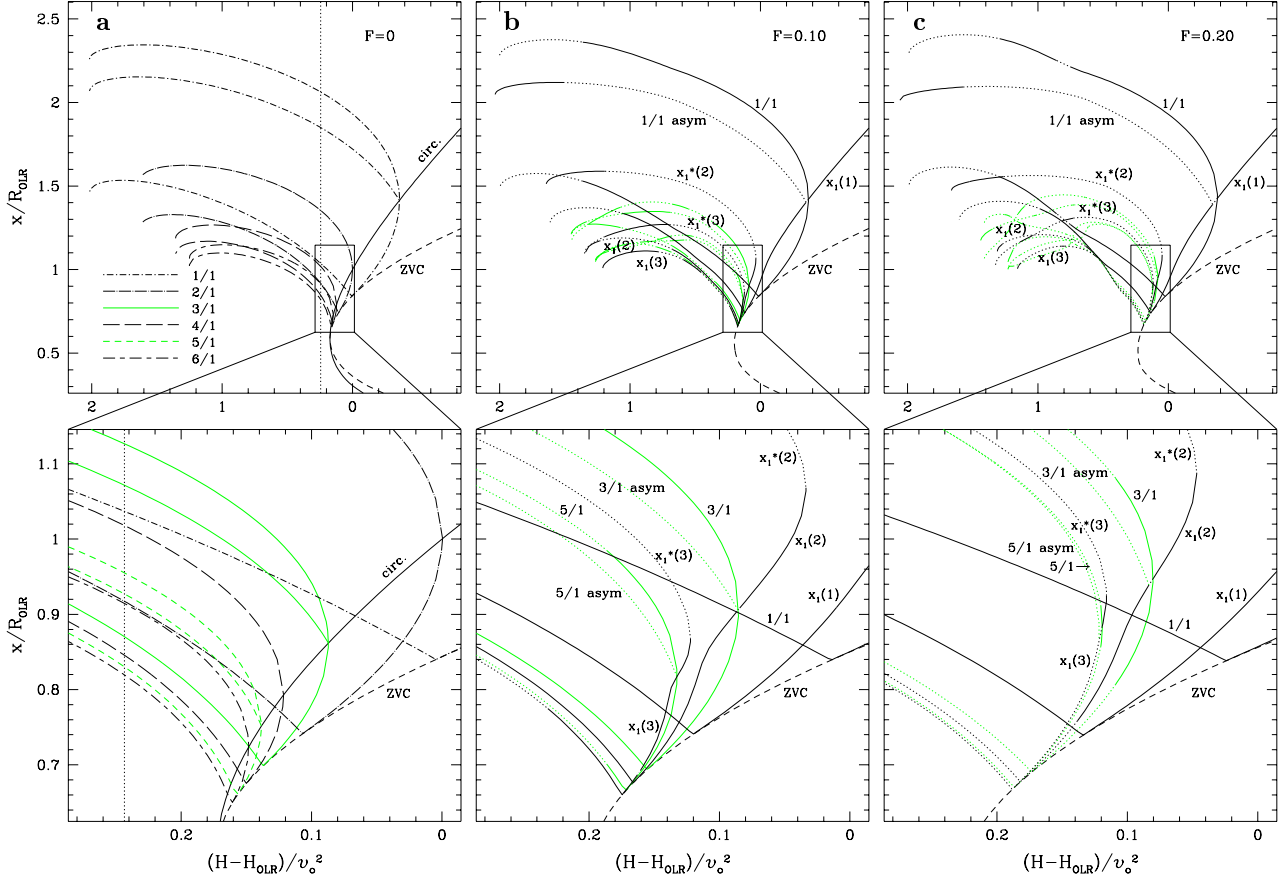}
\caption{\small {\bf a)} Characteristic diagram in the rotating axisymmetric
potential (Eq.~(\ref{axi})) for the circular orbit (circ.) and the lowest
order outer $n_R/1$ resonant orbits which are symmetric with respect to the
coordinate axes. The curves give the $x$-coordinate normalised by the OLR
radius of the orbits when they cross the \mbox{$x$-axis} with $\dot{y}>0$ as a
function of the Hamiltonian value relative to $H_{\hbox{\tiny OLR}}$, the
$H$-value of the circular orbit at the OLR, and normalised by $v_{\circ}^2$,
assuming that the potential rotates clockwise in the inertial frame. The
short-dashed line gives the zero velocity curve (ZVC) and the vertical dotted
line the $H$-value of the orbits sketched in Fig.~\ref{orb0}. The grey $3/1$
and $5/1$ resonance curves are shown only in the lower frame, which is a
magnification of the rectangular box in the upper frame.
{\bf b)} Corresponding characteristic diagram in the barred potential of
Eq.~(\ref{pot}) with a bar strength $F=0.10$. The major axis of the bar
coincides with the \mbox{$y$-axis}. The full and dotted parts of the curves
stand for stable and unstable orbits respectively. The orbit labels refer to
the nomenclature of Contopoulos \& Grosbol (\cite{CG}). The symmetric and
asymmetric $3/1$ and $5/1$ resonance curves are plotted with grey lines in
both the upper and lower frames, but labelled only in the lower one.
{\bf c)} Same as former diagram, but with $F=0.20$.}
\label{char}
\end{figure*}
\par The characteristic curves of these resonant orbits in the $H-x$ plane
(Fig.~\ref{char}a) all intersect the circular orbit curve (COC) at their point
of lowest $H$, corresponding to a {\it bifurcation}. For even resonances, four
branches emanate from the bifurcation, two from the COC and two from the
resonance curve. The resonance branches above (towards larger $x$) and below
the COC represent orbits with respectively apocentre and pericentre on the
\mbox{$x$-axis}. The lower branch always passes through the zero velocity
curve (ZVC), where the orbit becomes cuspy on the \mbox{$x$-axis} and then
develops loops at higher value of the Hamiltonian. For such loop orbits, the
\mbox{$x$-coordinate} of the characteristic curves does not trace the
pericentre but the place where the orbit self-intersect and $\dot{x}\neq 0$.
For odd resonances, the bifurcation has six branches: the two from the COC,
two for the resonant orbits with radial extrema on the \mbox{$x$-axis} and
which have properties similar to the former even resonance branches, and two
for the resonant orbits with those extrema on the \mbox{$y$-axis}. The two
latter branches have opposite $\dot{x}$ but degenerate into the same curve in
the $H-x$ characteristic diagram. All periodic orbits in the symmetry plane of
an axisymmetric potential are stable.
\par Figures~\ref{char}b and \ref{char}c show how the characteristic curves
are modified when the bar component with major axis on the \mbox{$y$-axis} is
added to the potential. The changes mainly occur at the bifurcations of the
axisymmetric case. The bifurcations of the even resonances become gaps, with
the right (low $H$) COC branch deviating into the lower resonance branch, and
the upper resonance branch into the left COC branch, giving rise to a sequence
of continuous orbit families. In the terminology introduced by Contopoulos \&
Grosbol (\cite{CG}), the outermost of these families is called $x_1(1)$, and
the other families are divided into an upper $x_1^*(i)$ and a lower $x_1(i)$
sub-family at or near the point of minimum $H$, where the stability of the
orbits appears to reverse. The six-branch bifurcations of the odd resonances
(see the $1/1$ and $3/1$ resonances in the figures) split into two pitchfork
bifurcations, one involving the resonant orbits symmetric with respect to the
bar minor axis, which are stable near the bifurcation, and the other the
resonant orbits non-symmetric relative to this axis, which are unstable near
the bifurcation and qualified as {\it asymmetric}. As a by-product of this
splitting, the segment of the $x_1(i)$ characteristic curve between the two
new bifurcations becomes unstable.
\begin{figure}
\includegraphics[width=8cm]{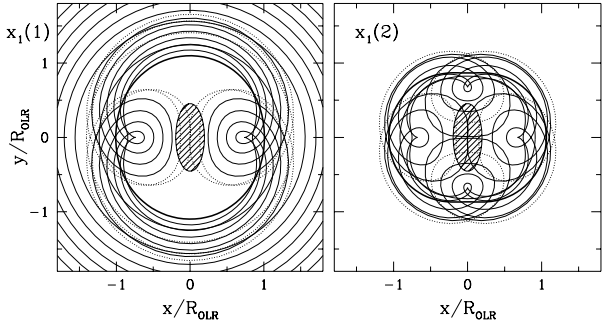}
\caption{\small Orbits of the $x_1(1)$ (left) and $x_1(2)$ (right) families in the
model with bar strength $F=0.10$, with the thick line representing the orbit
of lowest apocentre and of lowest Hamiltonian respectively. The shaded ellipse
sketches the bar. The full and dotted lines represent stable and unstable
orbits respectively. The nearly circular $x_1(1)$ orbits extend out to
infinity. Note the unstable $x_1(1)$ orbit at the $1/1$ resonance.}
\label{x1}
\end{figure}
\par At low eccentricity (i.e. small $H$), only those orbits with a pericentre
on the bar minor axis seem to be stable. This could be the influence of the
Lagrangian points $L_{\rm 4/5}$, which are stable fix points in the models. At
high eccentricity, all orbit families undergo a change of stability and the
characteristic curves in Fig.~\ref{char} are all interrupted at the point
where the minor axis loop of the orbits touches the centre $R=0$. The curves
actually go beyond this point, but the resonance number changes. For instance,
the $2/1$ orbits become $2/3$ orbits. When increasing the bar strength (from
Fig.~\ref{char}b to Fig.~\ref{char}c), the resonance gaps between successive
$x_1(i)$ curves and the separation between the pitchfork bifurcation pairs
increases. The characteristic curves also become more twisted and the fraction
of stable orbits decreases. In particular, almost all $x_1(3)$ orbits are
unstable for $F=0.20$. It should be noted that other authors (Athanassoula et
al. \cite{ABMP}; Sellwood \& Wilkinson \cite{SW}) report more complicated
characteristic curves for the $x_1(2)$ and higher order orbit families near
the ZVC in their models, probably as a consequence of an $m=4$ Fourier
component in the potential.
\par Some orbits of the above families are plotted in D2000, but obviously
missing all eccentric $x_1(1)$ and $x_1(2)$ orbits with loops on the bar minor
axis. A more complete set of orbits from these two families are given in
Fig.~\ref{x1}. These orbits are the most important even $n_R$ periodic orbits
because they are usually associated with the largest invariant curve islands
in surface of section maps. The orbits drawn with thick lines are examples of
the perpendicularly oriented orbits that replace the circular orbit near the
OLR radius and which have been proposed as a possible explanation for either
the Hyades and Sirius streams (Kalnajs \cite{K}) or the Hercules-LSR
bimodality (D2000) in the observed $u-v$ distribution. As we shall see in the
next sections, the stable eccentric $x_1(1)$ orbits also play an important
role in shaping the local velocity distribution. Regular orbits trapped around
them are indeed unlikely to be heavily populated by stars, but represent
forbidden phase space regions for chaotic orbits.
\par
\begin{figure*}[t!]
\includegraphics[width=17.5cm]{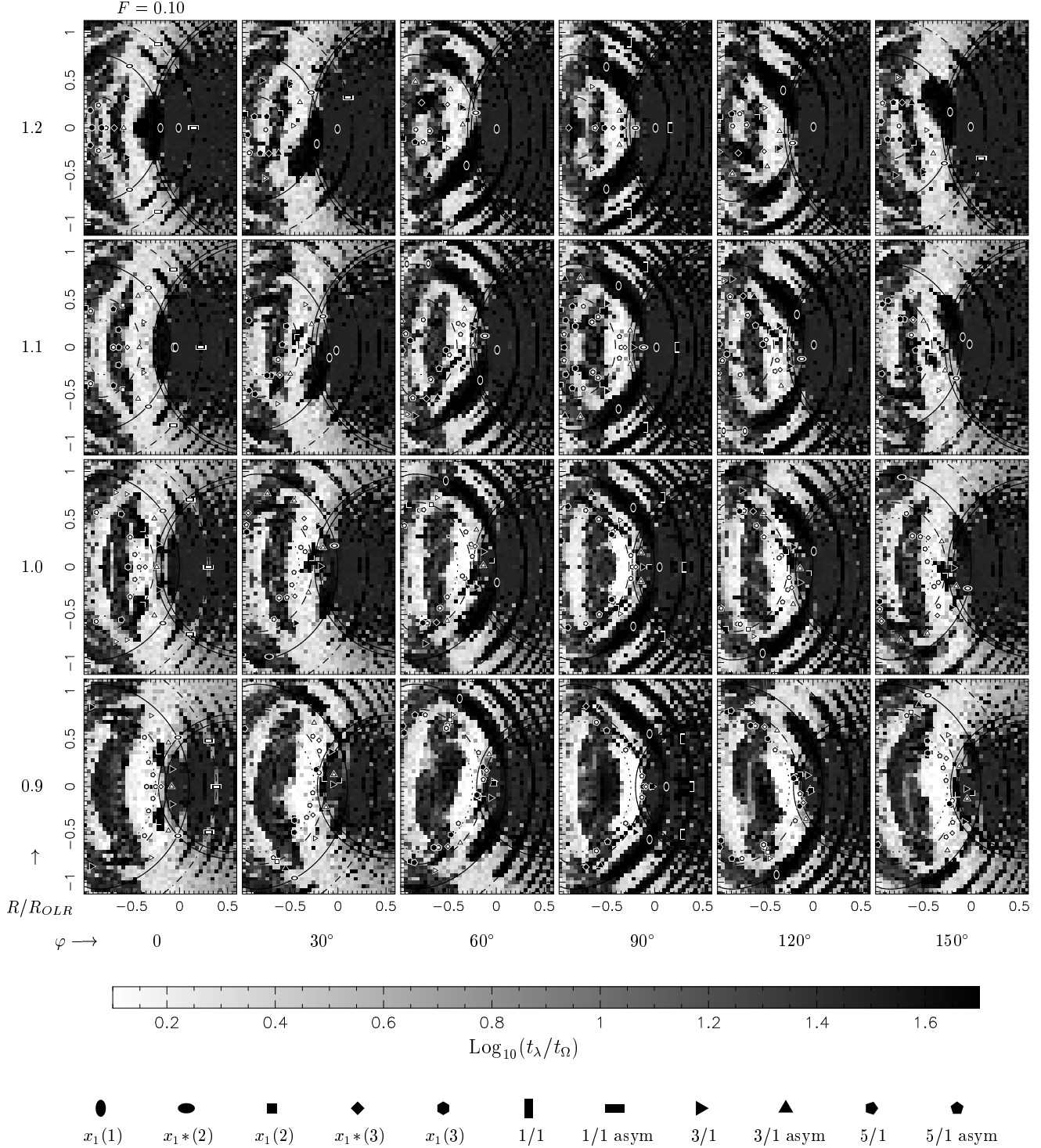}
\vspace*{0.1cm}
\caption{\small Liapunov divergence timescale of the orbits in the $u-v$ plane as a
function of position in real space, for a bar strength $F=0.10$. The
horizontal and vertical axes of each frame are $v/v_{\circ}$ and $u/v_{\circ}$
respectively, with $v$ positive in the direction of rotation and $u$ towards
the anti-centre, and the origin at the circular orbit of the axisymmetric part
of the potential $\Phi_{\circ}$. The timescales are in units of local circular
period~$t_{\Omega}$ in $\Phi_{\circ}$ and greyscale coded. The dark and white
regions respectively represent regular and chaotic orbits. The oval and
polygonal symbols indicate the positions of the periodic orbits, with a
different symbol for each orbit family. Full and empty symbols respectively
stand for stable and unstable orbits. The full lines open towards the right
(increasing $v$) are the $H_{12}$ and $H_{45}$ contours. The dash-dotted,
solid, dashed and dotted lines open towards the left respectively give the
locations of the outer $1/1$, $2/1$, $4/1$ and $6/1$ resonant orbits in
$\Phi_{\circ}$.}
\label{lia10}
\end{figure*}

\begin{figure*}[t!]
\includegraphics[width=17.5cm]{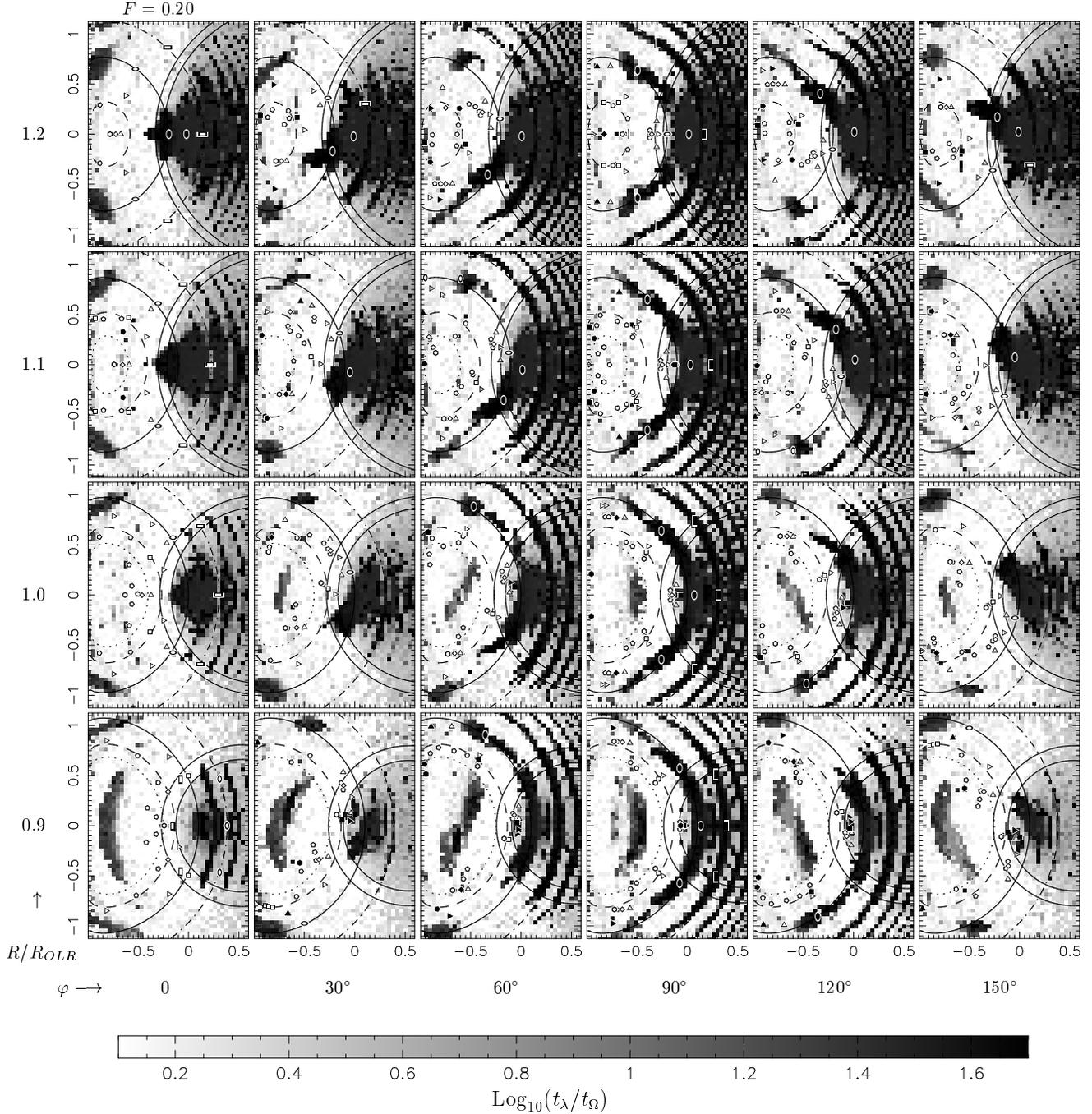}
\caption{\small As Fig.~\ref{lia10}, but for a bar strength $F=0.20$.}
\label{lia20}
\end{figure*}

 The space coverage of the orbit families can be determined from $x-y$
plots like in Fig.~\ref{x1}. For each position in real space, there will be an
(infinite) set of periodic orbits passing through. The velocity trace in
planar velocity space of the above described orbits, as well as the curves
delineating some of the main resonances in the underlying axisymmetric
potential\footnote{As pointed out in D2000, the figures 2 and 4 in
Fux (\cite{F3}) wrongly display the OLR as a contour of constant $H$. This
mistake is rectified here throughout the paper.}, are indicated in
Figs.~\ref{lia10} and~\ref{lia20} for various azimuthal angles~$\varphi$ and a
realistic range of galactocentric distances relative to the OLR, and for two
different bar strengths. Here the angle~$\varphi$ is measured from the bar
major axis and increases towards the direction opposite to galactic rotation,
i.e. coincides with the traditional in-plane inclination angle of the bar. All
space positions are reached by many of the considered orbit families, and
sometimes several traces are produced by orbits from the same family: for
instance, at $\varphi=90^{\circ}$, there is a large range of $R$ with three
traces from $x_1(1)$ orbits, mainly due to the loops on the \mbox{$x$-axis} of
the high-$H$ orbits. The traces of the $1/1$, $1/1$ asym, $x_1(1)$,
$x_1^*(2)$, $x_1(2)$ and $x_1^*(3)$ orbits with non-nearly vanishing
$|u|$-velocity all fall very close to the associated resonance curves,
indicating that the axisymmetric approximation used to compute these curves
works well for $n_R\leq 4$. Not all resonant periodic orbits, i.e. those with
traces on the resonance curves, are unstable. In particular, orbits on the
$2/1$ (OLR) and $1/1$ resonance curves are stable $x_1(1)$ and $1/1$ orbits at
$\varphi=90^{\circ}$ and unstable $x_1^*(2)$ and $1/1$ asym orbits at
$\varphi=0^{\circ}$, except for a $x_1^*(2)$ and a $x_1(1)$ orbit with $u=0$
for $R/R_{\hbox{\tiny OLR}}\geq 1.1$. Hence resonance regions of phase space
are not necessary unstable and depleted as asserted in D2000 for the OLR.
\par There are sometimes several orbits from the same family plotted very
close to each other, like for example the three unstable $x_1(2)$ orbits at
$R/R_{\hbox{\tiny OLR}}=0.9$, $\varphi =30^{\circ}$ and
$(v/v_{\circ},u/v_{\circ})\approx (-0.85,-0.80)$ for $F=0.20$. This happens
when the sequence of orbits within the family reverses its progression in the
$x-y$ plane towards a given direction and very close to the current space
position, causing an accumulation of orbits near this position with different
local velocities. One may also note that the continuous transitions between
some orbit families can cause periodic orbits from different families to have
almost identical traces, as for example the $x_1^*(2)$ and $x_1(2)$ near the
centre of the frame $R/R_{\hbox{\tiny OLR}}=1.1$ and $\varphi=90^{\circ}$ in
Fig.~\ref{lia20}.
\par\parbox{8.8cm}{\vspace*{0.4cm}}\par\parbox{8.8cm}{\vspace*{0.4cm}}
\par The rather large number of stable simple periodic orbits through each
space position and the numerous stellar streams observed in the Solar
neighbourhood may suggest that at least some of them are related to periodic
orbits, as anticipated in Kalnajs (\cite{K}). For instance, beside the idea of
the $x_1(1)$ and low-eccentricity $x_1(2)$ induced streams near the OLR
radius, the $\varphi=30^{\circ}$ frames in Fig.~\ref{lia10} betray interesting
stable eccentric $x_1(2)$ and $5/1$ asym orbits at
$R/R_{\hbox{\tiny OLR}}=1.2$ and $R/R_{\hbox{\tiny OLR}}=1.1$ respectively and
with $v/v_{\circ}\approx -0.6$ and $u/v_{\circ}\approx 0.05-0.15$, which fall
very close to the velocity of the young Arcturus stream (see
Table~\ref{stream}).

\section{Liapunov exponents}
%%%%%%%%%%%%%%%%%%%%%%%%%%%%
\label{liaexp}

The next step after the periodic orbit search is to determine the phase space
extent of the regular orbits trapped around the stable closed orbits and of
the chaotic orbits, which have no other integral than the Jacobi integral. The
Poincar\'e surface of section method is well suited to highlight the regular
and chaotic regions in phase space at constant value of the Hamiltonian, but
not at constant position in real space. A better tool for this purpose are the
Liapunov exponents, which also allow to quantify the degree of stochasticity
of the orbits. These exponents describe the mean exponential rate of
divergence of two trajectories initially close to each other in phase space
and are defined as:
\begin{equation}
\lambda(\vec{x}_{\circ},\Delta \vec{x}_{\circ})=
\lim_{\scriptsize \begin{array}{c} \Delta \vec{x}_{\circ}\rightarrow 0 \\
t\rightarrow \infty \end{array}}\frac{1}{t}
\ln{\frac{|\Delta \vec{x}(t)|}{|\Delta \vec{x}_{\circ}|}},
\label{lia}
\end{equation}
where $\vec{x}_{\circ}$ and $\Delta \vec{x}_{\circ}$ are the initial ($t=0$)
position and deviation, and $\Delta \vec{x}(t)$ is the deviation at time $t$.
Such limit is proven to exist and is finite for all bound orbits (Oseledec
\cite{O})\footnote{In a realistic disc galaxy, the chaotic orbits not confined
within corotation, and in particular the hot chaotic orbits, are not really
bound but their escaping timescale is much larger than the Liapunov divergence
timescale (see also Sect.~\ref{testpart}). This is especially true in the case
of our infinite mass logarithmic potential.}. The value of $\lambda$ generally
depends on the direction of the deviation $\Delta \vec{x}_{\circ}$ and, if $N$
is the dimension of phase space, one can show (Oseledec \cite{O}; Benettin et
al. \cite{BGS0}) that there exists in fact $N$ discrete exponents
$\lambda_1\geq \lambda_2\geq \ldots \geq \lambda_N$. However, the largest
exponent $\lambda_1$ is the most important because it results from almost all
deviations in $\Delta \vec{x}_{\circ}$-space (e.g. Udry \& Pfenniger
\cite{UP}), and therefore is the one found in practice when the initial
deviation is chosen randomly. Moreover, this exponent exclusively determines
the orbital stability: if $\lambda_1=0$, all exponents vanish and the orbit is
regular, and if $\lambda_1>0$, the orbit is chaotic and the amount of chaos
increases with $\lambda_1$.
\par The numerical computation of the Liapunov exponents faces some problems
related to the limits in Eq.~(\ref{lia}). First, the finite initial deviation
$\Delta \vec{x}_{\circ}$ may rapidly grow as large as the size of the orbits
themselves, especially for chaotic orbits, and thus $\Delta \vec{x}(t)$ must
be occasionally scaled down by a large factor. Noting $\Delta \vec{x}^0(t)$
the deviation before the first rescaling, this is done in a way similar to
Contopoulos \& Barbanis (\cite{CB}), by normalising $\Delta \vec{x}(t)$ to
$|\Delta \vec{x}_{\circ}|$:
\begin{equation}
\Delta \vec{x}^i(t_i)=\frac{|\Delta \vec{x}_{\circ}|}
{|\Delta \vec{x}^{i-1}(t_i)|}\Delta \vec{x}^{i-1}(t_i)
\label{scale}
\end{equation}
every time $t_i$ when
$|\Delta \vec{x}^{i-1}|>1.3\cdot 10^{-3}R_{\hbox{\tiny OLR}}$, so that after
$n$ iterations:
\begin{equation}
\ln{\frac{|\Delta \vec{x}(t)|}{|\Delta \vec{x}_{\circ}|}}=
\ln{\frac{|\Delta \vec{x}^n(t)|}{|\Delta \vec{x}_{\circ}|}}+
\sum_{i=1}^n \ln{\frac{|\Delta \vec{x}^{i-1}(t_i)|}{|\Delta \vec{x}_{\circ}|}}.
\label{iter}
\end{equation}
Another problem is that one cannot integrate orbits over an infinite time and
thus the time limit in Eq.~(\ref{lia}) must be replaced by the value at some
finite time, which has been set to about $3$ Hubble times when the models are
scaled to realistic physical units.
\par The Liapunov exponent $\lambda_1$ has been calculated on a Cartesian grid
of planar velocities for different positions of the observer, using the 2D
potential of Eq.~(\ref{pot}) and with an initial deviation of
$+1.3\cdot 10^{-11} R_{\hbox{\tiny OLR}}$ in the \mbox{$R$-coordinate}
(Figs.~\ref{lia10}, \ref{lia20} and~\ref{lia10R}). The results are presented
as a divergence timescale $t_{\lambda}\equiv 1/\lambda_1$ in units of local
circular period in the axisymmetric part of the potential, which provides a
more obvious and useful measure of stochasticity. Diagrams have been computed
for every $10^{\circ}$ in azimuth for $R/R_{\hbox{\tiny OLR}}=$ 0.9, 1.0,
1.05, 1.1, 1.2 and also at $\varphi=25^{\circ}$ over a larger radial range,
but those diagrams between our $30^{\circ}$ sampling and at
$R/R_{\hbox{\tiny OLR}}=1.05$ will not be shown here (and the same is true for
the velocity distributions of the next section). We shall refer to such
diagrams as ``Liapunov'' diagrams.
\par The first thing to notice in these diagrams is that the stable and
unstable periodic orbits fall within regular and chaotic regions respectively,
as expected. There are some apparent exceptions like the $1/1$~asym orbit at
$\varphi=0$ and large $R$ which must owe to the limited velocity resolution of
the diagrams ($=0.04v_{\circ}$; see Fux \cite{F5} for some higher resolution
Liapunov diagrams). The fraction of chaotic orbits also obviously increases
with bar strength. Furthermore, as a consequence of the four-fold symmetric
barred potential, the diagrams at $\varphi=0$ and $90^{\circ}$ are symmetric
with respect to $u=0$ and, more generally, diagrams at supplementary angles
are anti-symmetric to each other in $u$, i.e.
$t_{\lambda}(R,180^{\circ}-\varphi,v,u)=t_{\lambda}(R,\varphi,v,-u)$.
\par At $\varphi=90^{\circ}$ and for the radial range explored in
Figs.~\ref{lia10} and~\ref{lia20}, the $2/1$ and $1/1$ resonance curves, as
well as all other not highlighted resonances of the form $2/|n_{\phi}|$ with
integer $|n_{\phi}|$, are embedded in the middle of broad regular orbit arcs
separated by mainly chaotic regions which come closer to $u=0$ as the bar
strength increases. For $R/R_{\hbox{\tiny OLR}}\ga 1.1$ (and
$\varphi=90^{\circ}$), the regularity of the OLR arc gets destroyed near
$u=0$, leaving the place to an unstable $x_1^*(2)$ orbit. Between the dominant
regular orbit arcs, secondary regular arcs associated with resonances of the
form $4/|n_{\phi}|$ with odd integer $|n_{\phi}|$ can also be identified,
especially for $F=0.10$. This includes in particular the $4/1$ resonance arc
visible for $R/R_{\hbox{\tiny OLR}}\la 1.0$. In fact, the chaotic regions
between the broad resonance arcs are probably densely filled by tiny arcs of
higher order regular resonant orbits, but the filling factor must be very low.
\par At $\varphi=0$, the situation is reversed: the $2/|n_{\phi}|$ resonance
curves lie in chaotic regions at large $H$ which are spaced by regular orbit
arcs right between the resonances. At intermediate angles, the regular and
chaotic regions become offset from the resonance curves and the $u$-symmetry
breaks. In particular, for $\varphi \sim 30^{\circ}$ and
$R/R_{\hbox{\tiny OLR}}\ga 1.0$, i.e. realistic positions for the Sun, an
extreme case of asymmetry arises near the OLR: a prominent region of regular
orbits extends down to negative $u$, bounded roughly by the OLR curve on the
right and penetrating well inside the hot orbit zone, whereas the positive $u$
part of the OLR curve is surrounded by a wide chaotic region extending
somewhat inside the $H_{12}$ contour and coinciding very well with the $u-v$
location of the Hercules stream.
\begin{figure*}
\includegraphics[width=17.5cm]{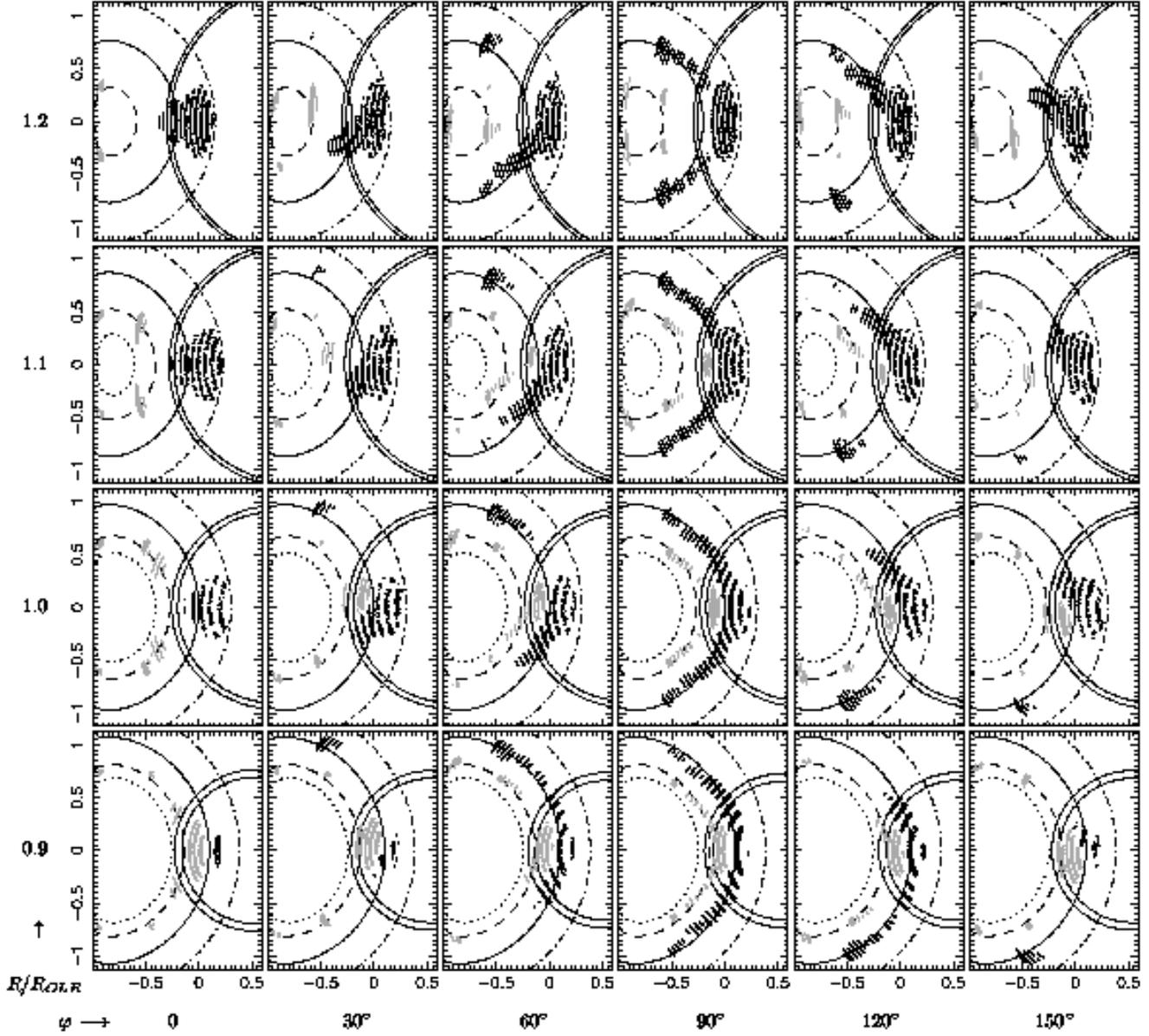}
\caption{\small Location in the $u-v$ plane of the regular orbits trapped around the
stable $x_1(1)$ periodic orbits for a bar strength $F=0.20$ (dark points) and
around the stable $x_1(2)$ orbits for $F=0.10$ (grey points). The
computational details are described in the text. The horizontal and vertical
axes of the frames and the different curves are as in Fig.~\ref{lia10}. The
$H$-contours are given for $F=0.10$.}
\vspace*{0.2cm}
\label{uvx1}
\end{figure*}
\par Figure~\ref{uvx1} shows the $u-v$ region occupied by the regular orbits
trapped around the stable $x_1(1)$ and $x_1(2)$ periodic orbits. To construct
this figure, we have first derived many surfaces of section at mainly constant
Hamiltonian interval $\Delta H=0.054v_{\circ}^2$ to locate the islands of
invariant curves around these periodic orbits. Then the orbits within each
islands of these maps have been sampled by 50 regularly spaced points along a
straight line across the central periodic orbit and within the outermost
invariant curve of the island. Finally, all the resulting initial conditions
are integrated for 20 rotations in the rotating frame and the velocities are
plotted when the orbits pass within a distance of $d_{\rm max}=R_{\circ}/80$
of the considered space positions. The striped appearance of the regular
$x_1(1)$ and $x_1(2)$ regions in the diagrams are due to the discrete
$H$-sampling and the broadening of the stripes to the finite value of
$d_{\circ}$ (not to an inaccurate orbital integration). The islands in the
surfaces of section generally contain sub-resonances which have been included.
The $x_1(1)$ islands however have been truncated at the $1/1$ resonance, so
that the part of the $x_1(1)$ region on the high-$v$ side of the $1/1$
resonance curve is not represented in the $u-v$ diagrams. The $x_1(1)$ regions
are derived for $F=0.20$ because the high level of chaos at this bar strength
makes it easier to distinguish the boundaries of these regions in the surfaces
of section, and the $x_1(2)$ regions for $F=0.10$ in order to emphasise the
more regular case where these regions are larger.
\par From this figure, it is obvious that the regular orbit arcs near the OLR,
and in particular the prominent regular region at $\varphi \sim 30^{\circ}$
and $R/R_{\hbox{\tiny OLR}}\ga 1.0$ discussed previously, are produced by the
regular orbits around the stable periodic orbits of the $x_1(1)$ family, with
an eccentricity increasing towards larger Hamiltonian values. The regular
regions associated with the stable $x_1(2)$ orbits can be viewed as divided
into two parts, one involving only low-eccentricity orbits and the other one
the higher eccentricity orbits falling close to the $4/1$ resonance curve. The
low-eccentricity orbit part exists for $R/R_{\hbox{\tiny OLR}}\la 1.1$ and
over an angle range around $\varphi=90^{\circ}$ increasing as $R$ decreases,
and is generally enclosed between the $H_{12}$ contour and the OLR curve.
Elsewhere, it is dissolved and only an unstable $x_1^*(2)$ orbit remains
(Fig.~\ref{lia10}). The higher eccentricity part connects the low-eccentricity
part at $R/R_{\hbox{\tiny OLR}}\approx 0.9$ and then progressively detach from
it as the $4/1$ resonance curve moves away from the $H_{12}$ contour at larger
$R$. In particular, for $R/R_{\hbox{\tiny OLR}}\ga 1.1$ and for $F=0.10$,
there are no regular \mbox{quasi-$x_1(2)$} orbits between the $H_{12}$ contour
and azimuthal velocities less than $v/v_{\circ}=-0.35$, whatever the angle
$\varphi$, and no low-eccentricity such orbits at all for
$\varphi \la 50^{\circ}$. Hence for this bar strength, the Hercules-like mode
found in D2000's simulations at
$R/R_{\hbox{\tiny OLR}}\approx 1.1$\footnote{More precisely at
$R_{\hbox{\tiny OLR}}/R=0.9$.} and $\varphi \sim 25$ cannot be related to such
\mbox{quasi-$x_1(2)$} orbits.
\begin{figure*}
\includegraphics[width=17.5cm]{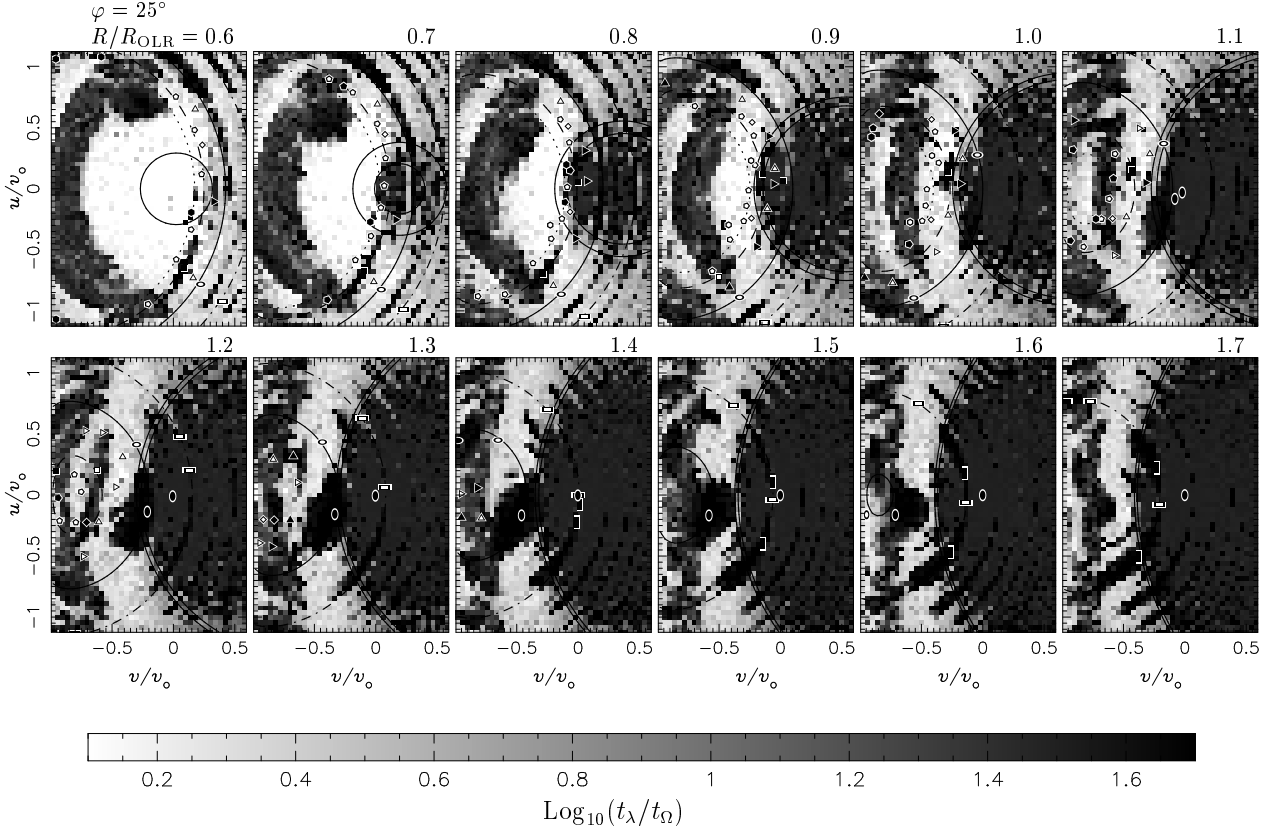}
\caption{\small As Fig.~\ref{lia10}, but for $\varphi=25^{\circ}$ and
$0.6\leq R/R_{\hbox{\tiny OLR}}\leq 1.7$.}
\label{lia10R}
\end{figure*}
\par Figure~\ref{lia10R} provides Liapunov diagrams over a larger radial range
at $\varphi=25^{\circ}$ and for $F=0.10$. These diagrams clearly show that the
$H_{12}$ contour marks the average transition between regular and chaotic
motion. Hence disc orbits are essentially regular and hot orbits essentially
chaotic. The diagrams in Fig.~\ref{lia10R} also nicely illustrate the
dependence of the $H_{12}$ and $H_{45}$ contours with galactocentric distance
discussed in Sect.~\ref{hampot}, and show how the velocity spacing between
these contours decreases with increasing $R$.
\par Martinet \& Raboud (\cite{MR}) have computed a diagram $\Delta(v,u)$
representing the relative pericentre deviation between planar orbits
integrated in a barred \mbox{$N$-body} model and the corresponding orbits in
the underlying axisymmetric potential, starting from a realistic space
position of the Sun. Their diagram correlates well with our Liapunov diagrams
in the sense that the larger values of $\Delta$ coincide with shorter Liapunov
timescales. In particular, a clear jump of $\Delta$ occurs at
$H\approx H_{\rm 12}$, with the high-$H$ side displaying much larger
pericentre deviations on the average, and there is also a tail of small
$\Delta$-values at negative $u$ extending inside the hot orbit region.
\par It is worth mentioning that in a two-dimensional axisymmetric potential,
all orbits are regular, i.e. have vanishing $\lambda_1$. Hence the chaotic
regions discussed here are all produced by the influence of the bar alone.
Also, the divergence timescale $t_{\lambda}$ in the chaotic regions may be as
low as a few orbital times. This property may have important consequences on
the early evolution of the distribution function in barred galaxies, as we
shall see in the following section.

\section{Phase space crowding}
%%%%%%%%%%%%%%%%%%%%%%%%%%%%%%
\label{testpart}

Given the orbital structure of phase space, we now want to know how nature
populates the available orbits. This is done resorting to test particle
simulations with the following two integral initial distribution function
\pagebreak (Dehnen \cite{D3}):
\begin{equation}
f_{\circ}(E,L_z)\!=\!\frac{2\Omega(R_{\rm c})}{\kappa(R_{\rm c})}
\frac{\tilde{\Sigma}(R_{\rm c})}{2\pi \tilde{\sigma}(R_{\rm c})}
\exp{\!\left[\Omega(R_{\rm c})\frac{L_z\!-\!L_{\rm c}(E)}
{\tilde{\sigma}^2(R_{\rm c})}\right]},
\label{init}
\end{equation}
where $E$ and $L_z$ are respectively the energy and the $z$-component of the
angular momentum, $R_{\rm c}(E)$ and $L_{\rm c}(E)$ the radius and angular
momentum of the circular orbit with energy~$E$, $\Omega$ and $\kappa$ the
circular and epicycle frequencies, and $\tilde{\Sigma}(R)$ and
$\tilde{\sigma}(R)$ the approximate surface density and radial velocity
dispersion profiles. This is a modified Shu (\cite{S}) distribution where the
radius of the guiding centre $R_{\rm c}$ is deduced from the energy instead of
the angular momentum, with the main advantage that the density function
extends smoothly towards negative $L_z$. Adopting
$\tilde{\Sigma}(R)\sim \exp{(-R/h_R)}$ and $\tilde{\sigma}^2(R)=
\tilde{\sigma}_{\circ}^2\exp{(-(R-R_{\circ})/h_{\sigma})}$, where $R_{\circ}$
is the galactocentric distance of the Sun and $\tilde{\sigma}_{\circ}$ the
approximate velocity dispersion at that distance, the initial distribution
function has three free parameters, which unless otherwise specified are set
to $h_R/R_{\circ}=0.33$, $\tilde{\sigma}_{\circ}/v_{\circ}=0.2$ and
$h_{\sigma}/R_{\circ}=1.0$. This results in exactly the same initial
conditions as for the default flat rotation curve case in D2000. In the
beginning of the simulations, the non-axisymmetric part of the potential is
gradually switched on from no contribution at $t=0$ to its full value at
$t=2t_{\rm b}$, where $t_{\rm b}=2\pi/\Omega_{\rm P}$ is the rotation period
of the bar, exactly the same way as in D2000 for the simulations with default
integration time.
\par For the time integration, D2000 uses a subtle backward integration
technique based on the conservation of the phase space density along the
orbits in collisionless systems. The idea is to integrate back in time until
$t=0$ the phase space points on a Cartesian grid of $u-v$ velocities at a
given space position $(R_{\circ},\varphi)$ and time $t_{\rm end}$, to
determine the energy $E$ and angular momentum $L_z$ of the originating orbits
in the initial axisymmetric potential, and from these infer
$f(t_{\rm end},R_{\circ},\varphi,v,u)=f_{\circ}(E,L_z)$. The advantages of
this technique is that only the orbits strictly necessary to derive the
evolved local velocity distributions need to be computed, and that these
velocity distributions are not affected by Poisson noise. Moreover, a unique
simulation suffices to get $u-v$ distributions for different initial
conditions, because $f_{\circ}$ comes in only {\it after} the orbit
integration.
\par
\begin{figure}
\includegraphics[width=8cm]{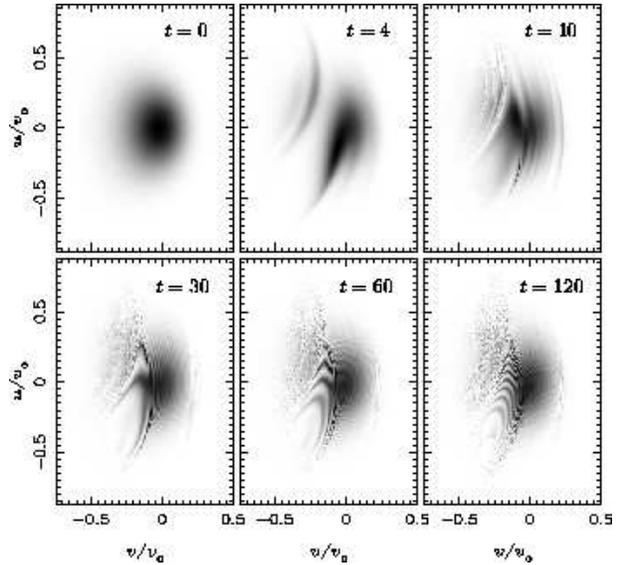}
\caption{\small Time evolution of the velocity distribution in the $u-v$ plane at
$R_{\circ}/R_{\hbox{\tiny OLR}}=1.1$ and $\varphi=25^{\circ}$ using the
backward integration technique. The bar strength is $F=0.10$ and the time is
given in units of the bar rotation period $t_{\rm b}$.}
\label{phasemix}
\end{figure}

 Unfortunately, the backward integration technique faces two major
problems illustrated in Fig.~\ref{phasemix}, which shows the long term
evolution of the planar velocity distribution at
$R_{\circ}/R_{\hbox{\tiny OLR}}=1.1$ and $\varphi=25^{\circ}$ using this
technique. The integration time in D2000 ranges from $4$ bar rotation for most
simulations, corresponding to only $\sim 2$ orbital periods at $R=R_{\circ}$
in the inertial frame, up to $8$ bar rotation for some cases, but always
matching the growth time of the bar to half the total integration time. The
frame at $t=4t_{\rm b}$ is similar to his results, revealing a clear bimodal
distribution. However, at $t=10t_{\rm b}$, the valley between the two modes
becomes heavily populated, destroying the bimodality, and at later times,
incurved waves appear in this valley with a spacing between the maxima
decreasing with time. This is a typical signature of ongoing phase mixing in a
regular region of phase space, which here corresponds to the eccentric orbit
part of the \mbox{quasi-$x_1(1)$} region according to the previous section. A
very similar phenomenon, with similarly incurved waves, can also be barely
recognised near the $1/1$ resonance. Between the $2/1$ and $1/1$ resonances,
phase mixing operates on a shorter timescale and the orientation of the wave
fronts seems to change from nearly-vertical at $t=30t_{\rm b}$ to
nearly-horizontal at $t=120t_{\rm b}$. The backward integration technique in
fact yields the fine-grained distribution function, which never smoothes out
on sufficiently small scales, whereas the physical one to compare with the
observations is the coarse-grained distribution. The second problem is related
to chaos: at $t\ga 30t_{\rm b}$, the $u-v$ distribution becomes noisy in the
chaotic regions because the phase space points integrated backwards from these
regions sample the initial distribution function more randomly. Hence much
longer integration times than in D2000 are required to obtain
quasi-equilibrium (coarse-grained) distribution functions and to properly take
into account the effect of chaos (remember that the divergence timescales for
chaotic orbits is of the order of several orbital periods), and one cannot
escape the fate of smoothing the fine-grained distribution.
\begin{figure*}
\vspace{16.59cm}
\caption{\small Velocity distribution in the $u-v$ plane as a function of space
position in the test particle simulations with a bar strength $F=0.10$. The
distributions are time averages within the interval from $25$ to $35$ bar
rotations after the beginning of the simulations. The velocity contours are as
in Fig.~\ref{obs}, whereas the axis labels, the Hamiltonian contours, the
resonance curves and the periodic orbits are as in Fig.~\ref{lia10}.}
\label{pop10}
\end{figure*}
\par Therefore, most of the test particle simulations in this paper were done
by simple forward integration. The initial phase space density is sampled by
$N$ particles which are then integrated forward in time, and the $u-v$
diagrams at space position $(R_{\circ},\varphi)$ are constructed from all the
particles within a distance $d_{\rm max}=R_{\circ}/80$ from that position
(corresponding to $d_{\rm max}=100$~pc for $R_{\circ}=8$~kpc), using a
Cartesian velocity binning with a bin size of $0.005v_{\circ}$. To increase
the particle statistics, the $u-v$ distributions are averaged over 10 bar
rotations and then smoothed within a square of $11\times 11$ bins. The time
average, which is hardly possible in the backward integration technique, is
also very convenient to reduce the phase mixing problem, as the contribution
of each particle to a $u-v$ distribution becomes proportional to the time
the particle spends within the volume where the distribution is computed.
With the forward integration technique, the time evolution of the velocity
distribution can be followed within a unique simulation. The test particle
simulations of this paper all have $N=10^6$ and the velocity distributions
have been derived within two time intervals, $25\leq t/t_{\rm b}\leq 35$ and
$55\leq t/t_{\rm b}\leq 65$. Since the distance parameters in $f_{\circ}$
scale as $R_{\circ}$ and not $R_{\hbox{\tiny OLR}}$, the results at different
$R_{\circ}/R_{\hbox{\tiny OLR}}$ require distinct simulations.
\begin{figure*}
\vspace{16.2cm}
\caption{\small Same as Fig.~\ref{pop10}, but for a bar strength $F=0.20$ and a time
average over the interval $55\leq t/t_{\rm b}\leq 65$.}
\label{pop20}
\end{figure*}
\par Figure~\ref{pop10} shows the $u-v$ distributions at various space
positions averaged over the time interval $25\leq t/t_{\rm b}\leq 35$ for a
bar strength $F=0.10$. As for the Liapunov diagrams, the distributions are
obviously symmetric with respect to $u=0$ for $\varphi=0$ and
$\varphi=90^{\circ}$ and the distributions at same radius but supplementary
angles are anti-symmetric to each other in~$u$. This is clearly not the case
in the simulations of D2000 (see his figure~2, where $F=0.089$), providing a
further argument that these have not achieved a quasi-stationary regime.
Moreover, the traces of the stable $x_1(1)$ periodic orbits away from the
$2/1$ resonance curve lie closer to the high angular momentum peak of the
velocity distributions.
 For this bar strength and the adopted values of the
parameters in the initial distribution function, there is also no clear
bimodality with a deep separation valley at
$R_{\circ}/R_{\hbox{\tiny OLR}}\ga 1.1$ and near $\varphi=30^{\circ}$,
although a clear density excess remains at low $v$ and positive
$u$\footnote{Doubling the average integration time can reinforce the
bimodality, as shown in Fig.~\ref{best}a.}. However, all space positions where
a low-eccentricity regular $x_1(2)$ region exists (see former section and
Fig.~\ref{uvx1}) present a nice bimodality, with the low angular momentum mode
coinciding very well with that region and always peaking inside the $H_{12}$
contour.
\par While the traces of the non-resonant $x_1(1)$ and $x_1(2)$ orbits are
generally embedded within their associated quasi-periodic orbit modes, they do
not necessarily exactly coincide with the peak of these modes, especially in
the $x_1(2)$ case (e.g. $R_{\circ}/R_{\hbox{\tiny OLR}}=1.0$ and
$\varphi=60^{\circ}-120^{\circ}$ in Fig.~\ref{pop10}). Furthermore, since the
quasi-periodic orbits cover a larger space area than the periodic orbits
themselves, \mbox{quasi-$x_1(1)$} and \mbox{quasi-$x_1(2)$} modes may occur
even at positions where no $x_1(1)$ or $x_1(2)$ orbit are passing through.
\par Increasing the bar strength (Fig.~\ref{pop20}) provides a better
understanding of how the velocity distributions are affected by chaos. Now,
there appears to be an obvious second source producing a low angular momentum
mode, which adds to the \mbox{quasi-$x_1(2)$} orbit flow at the space regions
reached by these orbits, and acts alone elsewhere, as for instance at
$R_{\circ}/R_{\hbox{\tiny OLR}}=1.1$ and $\varphi=30^{\circ}$. The overdensity
in velocity space generated by this second source correlates very well with
highly stochastic regions in the Liapunov diagrams (see Fig.~\ref{lia20}) and
seems to always peak outside the $H_{12}$ contour. At $\varphi =90^{\circ}$,
the overdensity culminates at $u\approx 0$ and is enclosed by the regular arc
of eccentric \mbox{quasi-$x_1(1)$} orbits. At $\varphi=0$, these
\mbox{quasi-$x_1(1)$} orbits occupy the $u=0$ region and chaotic overdensities
happen symmetrically on both positive and negative $u$ sides of this region.
At $\varphi\sim 30^{\circ}$, the \mbox{quasi-$x_1(1)$} region is located at
negative $u$ and there is a large chaotic overdensity at positive $u$.
\par These properties result from the decoupling between the regular and the
chaotic regions of phase space. Since chaotic orbits cannot visit the regions
of regular motion and, vice versa, regular orbits avoid the chaotic regions,
the distribution function in each of these regions evolves in a completely
independent way. In the regular regions, it recovers roughly the initial
distribution after phase mixing, whereas in the chaotic regions, it is
substantially modified through a process known as {\it chaotic mixing} and
which operates on the Liapunov divergence timescale (e.g. Kandrup \cite{HEK}):
the particles on chaotic orbits quickly disperse within the easily accessible
phase space regions, i.e. not impeded by cantori or an Arnold web, and
converge towards a uniform population of these regions. The dominant
manifestation of chaotic mixing is a net migration of particles from the inner
to the outer space regions. For instance, in the simulations with
$R_{\circ}/R_{\hbox{\tiny OLR}}=1.1$, the scale length of the radial particle
distribution, which remains very close to an exponential in the range
$0.5\la R/R_{\circ}\la 1.25$, increases by $\sim 30$\% for $F=0.10$ and by
$\sim 90$\% for $F=0.20$ within this radial range. This migration is
particularly marked for particles on hot chaotic orbits because the region
inside corotation initially represents a large reservoir of such particles and
because these particles can freely pass over corotation. As a consequence, in
the explored $R_{\circ}/R_{\hbox{\tiny OLR}}$ range, the chaotic regions in
the $u-v$ diagrams are more heavily crowded than the regular regions at
$H\ga H_{12}$, therefore corresponding to local overdensities.
\par At first glance, there seems to be a rather continuous transition from
the \mbox{quasi-$x_1(2)$} orbit mode to the main chaotic orbit mode when
moving across the OLR radius towards increasing $R_{\circ}$, with always a
single effective peak showing up and with the involved \mbox{quasi-$x_1(2)$}
region progressively dissolving in the chaotic one. But in some cases, the two
mode-generating sources really contribute to distinct peaks in the velocity
distribution (see Fig.~\ref{best}d for an example).
\par The process of chaotic mixing leads to velocity distribution contours
which are parallel to the contours of constant $H$ in the chaotic regions
(e.g. Fig.~\ref{pop20}). This property is also consistent with Jean's theorem
stating that the distribution function in a steady-state system depends only
on the integral of motions. The only integral for the chaotic orbits in the
present 2D barred models is the value of the Hamiltonian, hence the
distribution function and therefore the corresponding velocity distributions
at fixed space position should be a function of only $H$ in the chaotic
regions. It should be noted that the Jeans theorem does not strictly apply to
the hot and disc chaotic orbits. Indeed, these orbits are not energetically
bound (in terms of~$H$) and thus the phase space density around such orbits
and within the finite space volume of the galaxy should decrease with time,
conflicting with the steady-state assumption of the theorem. However, the
escaping timescale of these chaotic orbits, which is essentially controlled by
Arnold diffusion across the confining cantori, is much longer than the
Liapunov divergence timescale and even the Hubble time, and thus the density
in the phase space regions covered by these orbits can be considered as almost
constant (see also Kaufmann \& Contopoulos \cite{KC} and references therein).
\par Secondary chaotic orbit overdensities also occur between the $x_1(1)$ and
the $1/1$ regular arcs, especially at $R_{\circ}/R_{\hbox{\tiny OLR}}\geq 1.0$
and $30^{\circ}\leq \varphi \leq 150^{\circ}$ (Fig.~\ref{pop20}). These
secondary overdensities and the above described main overdensities connect to
each other in phase space, i.e. are traced by the same orbits. Hence it is
also expected that the $u-v$ density at constant $H$-value is the same for all
overdensities. This is only roughly the case in Fig.~\ref{pop20}, probably
because the smoothing of the diagrams lowers the peaks of the tiny secondary
overdensities relative to the broader main overdensities. Small chaotic
overdensities may sometimes even be noticed beyond the $1/1$ resonance curve.
However, at high angular momentum, the hot orbits spend most of their time in
the outer galaxy, far away from the influence of the bar, and thus become more
regular (the energy and angular momentum are more nearly conserved). The
eccentric $x_1(2)$ regular regions can also represent density depressions
between chaotic regions in the velocity distributions, as can be marginally
inferred for example from the $R_{\circ}/R_{\hbox{\tiny OLR}}=0.9$ and
$\varphi =90^{\circ}$ frame in Figs.~\ref{uvx1} and~\ref{pop10}.
\par The valley between the main high-$L_z$ velocity mode (or LSR mode after
Dehnen) and the main chaotic orbit mode is generally close to the $H_{12}$
contour, reflecting the decline of the hot orbit population as
$H\rightarrow H_{12}$. Such a valley should in principle also exist between
the LSR mode and the secondary chaotic overdensities (see for instance
$R_{\circ}/R_{\hbox{\tiny OLR}}=1.1$ and $\varphi=30^{\circ}$ in
Fig.~\ref{pop20}). The main chaotic orbit mode also seems to always peak
between the $H_{12}$ and $H_{45}$ contours in Fig.~\ref{pop20}, but this is
not true for all our test particle simulations, as demonstrated by the
$h_R/R_{\circ}=0.2$ frame in Fig.~\ref{param} and Fig.~\ref{best}a. However,
this property might be more generic for self-consistent models (see
Sect.~\ref{nbody}). For some not fully understood reasons, the symmetry
properties mentioned previously for the case $F=0.10$ are somewhat less
evident for $F=0.20$, despite the longer integration time.
\par D2000 attributes the valley between the main LSR mode and the
Hercules-like mode to stars scattered off the OLR, in the sense that the
resonance generates chaotic orbits. In particular, he claims that the unstable
$x_1^*(2)$ orbit falls exactly between the two modes. This is not quite
correct for a stochastically induced Hercules-like mode, as is best
illustrated by Fig.~\ref{pop20}: for $R_{\circ}/R_{\hbox{\tiny OLR}}\ga 1.0$
and $\varphi\sim 30^{\circ}$, the $u>0$ part of the $2/1$ resonance curve
passes through the Hercules-like mode and the $x_1^*(2)$ orbit clearly lies
within the mode at $R_{\circ}/R_{\hbox{\tiny OLR}}=1.2$, and the low density
region below this mode is due to {\it regular} resonant orbits. D2000 also
claims that in his simulations the extension of the LSR mode to $u<0$ at
$v\approx -0.1v_{\circ}$ is caused by the elongation of the (presumably
\mbox{quasi-$x_1(1)$}) orbits near the OLR. Our results indicate that at least
the final part of this extension, corresponding to the secondary
overdensities, is produced by chaotic orbits. Such an extension exits in the
observations, but is only significant down to heliocentric
$u\approx -60$~km\,s$^{-1}$.
\par The velocity distributions at the two different mean integration times
$<\!t\!>=30t_{\rm b}$ and $60t_{\rm b}$ reveal some secular evolution. As
$<\!t\!>$ increases, the crowding contrast between the regular and chaotic
regions becomes more evident, with denser high-$H$ chaotic regions and deeper
regular region valleys. The quasi-$x_1(1)$ mode squeezes towards its high
angular momentum side for $R_{\circ}/R_{\hbox{\tiny OLR}}\ga 1.1$ and,
especially in the case $F=0.10$, the peak of the \mbox{quasi-$x_1(2)$} mode
moves closer to the $H_{12}$ contour for $R_{\circ}=R_{\hbox{\tiny OLR}}$,
betraying a longer phase mixing timescale near this resonance\footnote{This
could be a consequence of the fact that the linear radial oscillation
frequency of the quasi-periodic orbits around the least eccentric stable
$x_1(2)$ orbits near the OLR radius, i.e. the non-axisymmetric analogue of the
epicycle frequency for these periodic orbits, is close to the radial frequency
$w_R$ of the $x_1(2)$ orbits themselves. A similar argument may also hold for
the slow phase mixing noticed in Fig.~\ref{phasemix} within the resonant part
of the quasi-$x_1(1)$ region.}.
\par In a real galaxy, the presence of mass concentrations like giant
molecular clouds and of transient spiral arms will prevent the strict
conservation of the Jacobi integral and cause the stars to diffuse from the
regular regions to the chaotic regions of phase space and vice versa (see also
Sect.~\ref{nbody}). The chaotic regions should be very efficient in heating
galactic discs and the communication between the two kind of regions may even
allow to heat regular regions. The quantification of this phenomenon might be
an interesting problem to study, but is beyond the scope of this paper.
\par Finally, Figs.~\ref{pop10} and~\ref{pop20} also suggest that with
increasing bar strength, the velocity dispersion ratio $\sigma_{v}/\sigma_{u}$
decreases and, as reported by D2000, the $u$-range and in particular the mean
$u$-velocity of the main chaotic overdensity become larger.

\section{Changing the initial conditions}
%%%%%%%%%%%%%%%%%%%%%%%%%%%%%%%%%%%%%%%%%
\label{incond}

\begin{figure}
\includegraphics[width=8cm]{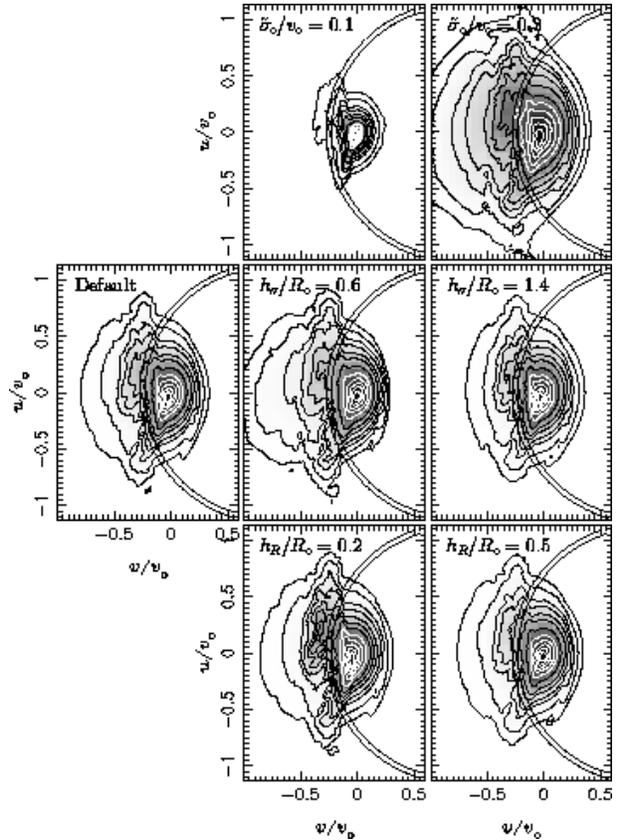}
\caption{\small Velocity distribution in the $u-v$ plane after $120$ bar rotations as
a function of initial conditions and using the backward integration technique.
The space position is $R_{\circ}/R_{\hbox{\tiny OLR}}=1.1$ and
$\varphi=25^{\circ}$ and the bar strength $F=0.10$. The left frame shows the
result for the default values of the parameters, i.e.
$\tilde{\sigma}_{\circ}/v_{\circ}=0.2$, $h_{\sigma}/R_{\circ}=1$ and
$h_R/R_{\circ}=0.33$, and the other frames the results when changing only one
parameter at a time to the indicated value. The velocity contours are as in
Fig.~\ref{obs} and the circular arcs represent the $H_{12}$ and $H_{45}$
contours.}
\label{param}
\end{figure}

How can the initial conditions be changed in order to increase the population
of the hot chaotic orbits, and in particular the density in the Hercules-like
chaotic overdensity~? Figure~\ref{param} shows how the three parameters of the
initial axisymmetric distribution function (Eq.~(\ref{init})) individually
affect the final velocity distribution for a realistic position of the Sun and
$F=0.10$, using exceptionally the backward integration technique which, as
mentioned in Sect.~\ref{testpart}, is very convenient for this purpose. A~long
time integration is adopted to reduce the phase mixing problem and the $u-v$
distributions are smoothed the same way as the previous ones based on the
direct integration method. A~comparison of the default parameter velocity
distribution with the corresponding distribution (at $\varphi=30^{\circ}$) in
Fig.~\ref{pop10} indicates that both integration techniques give very similar
results.
\par Increasing the overall initial velocity dispersion (top frames in
Fig.~\ref{param}) yields a larger final
velocity dispersion. Hence in this kind of simulations the particles remember
the initial conditions and the action of the bar is unable to completely erase
them. Also, the $u-v$ density in the Hercules-like stream is enhanced relative
to the density within the main velocity mode. This is because the larger
velocity dispersion lowers the peak of the latter mode, and because a larger
$\tilde{\sigma}_{\circ}$ increases the average Hamiltonian value of the
particles and thus the fraction of hot chaotic particles. Reducing the initial
velocity dispersion scale length keeping the same velocity dispersion at
$R_{\circ}$ (middle frames) renders the inner regions hotter and hence mainly
increases the density of the velocity distribution at low angular momentum. To
increase the relative fraction of stars in the Hercules-like stream, the most
efficient way seems to start with a smaller disc scale length (bottom frames).
This represents a higher initial space density of the disc in the inner
regions where the particles have larger $H$-values on the average and thus
again a larger fraction of hot chaotic particles which will be spread over the
whole disc by the barred potential.
\begin{figure}
\includegraphics[width=8cm]{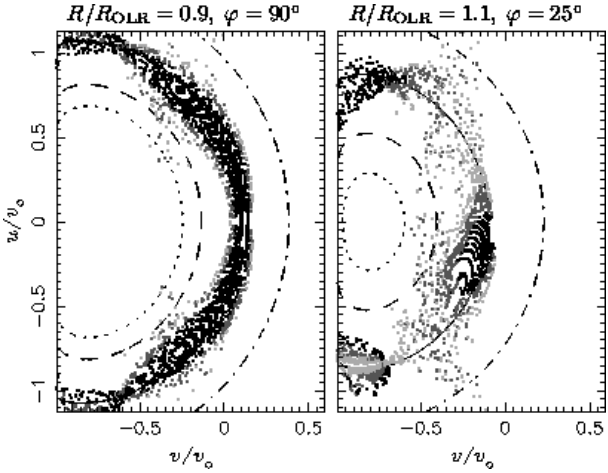}
\caption{\small For two different space positions and after $120$ bar rotations,
traces in the $u-v$ plane of the trajectories which were in OLR with the
unperturbed rotating potential at $t=0$ (bar strength $F=0.10$). The points
are colour-coded according to the angle $\alpha$ between the line joining the
apocentres of the initial resonant orbit and the major axis of the vanishing
bar potential: black if $|\alpha|<45^{\circ}$, light grey if
$|\alpha|>70^{\circ}$, and dark grey for intermediate angles.}
\label{reson}
\end{figure}
\par By changing the initial conditions, it is therefore possible to achieve a
more pronounced main chaotic overdensity than in the results based on the
default parameters and also to match more precisely the observed velocity
dispersions in the Solar neighbourhood.

\section{Resonances and stochasticity}
%%%%%%%%%%%%%%%%%%%%%%%%%%%%%%%%%%%%%%
\label{restock}

A worthful exercise is to determine what happens to the periodic orbits which
are initially in (outer) $2/1$ resonance before the growth of the bar. This is
illustrated in Fig.~\ref{reson}, which highlights for two different space
positions at $t=120t_{\rm b}$ the points in the $u-v$ plane corresponding to
trajectories which were on such orbits at $t=0$. The diagrams are built by
integrating backwards the trajectories passing through a Cartesian $u-v$ grid
and marking all the points on this grid originating from initial orbits with
$(\omega_{\phi}+2\omega_R-\Omega_{\rm P})/\Omega_{\rm P}<10^{-2}$. The
darkness of the points reflects the angle $\alpha$ between the major axis of
the initial resonant orbit and the major axis of the then vanishing bar
potential, with darker points standing for smaller angles, i.e. resonant
orbits with apocentre closer to the bar major axis.
\par At $\varphi=90^{\circ}$ and $R/R_{\hbox{\tiny OLR}}=0.9$, there is a wide
region of regular orbits around most of the $2/1$ resonance curve (see
Fig.~\ref{lia10}). The trajectories with small $\alpha$'s clearly fall in the
inner part of this region, while those with larger $\alpha$'s appear rather on
its edge and are spread within the neighbouring chaotic regions. In addition
to the regular versus chaotic phase space decoupling, the fact that the OLR
curve is associated with a valley in the planar velocity distribution (see
Fig.~\ref{pop10}) is also because the phase space density in the chaotic
regions bounding the regular orbit arc around this curve is forced to be a
function of $H$ only, hence lowering the density on the low-$v$ side and
increasing it on the other side relative to the initial densities. At
$\varphi=25^{\circ}$ and $R/R_{\hbox{\tiny OLR}}=1.1$, the part of the $2/1$
resonance curve within the $u>0$ chaotic region has no nearby small-$\alpha$
points and is embedded in a broad cloud of high-$\alpha$ points.
\par Hence, the initial $2/1$ resonant orbits more nearly aligned with the bar
major axis end into regular regions, while only those more nearly
perpendicular to this axis are scattered into chaotic regions. This is
consistent with the stability properties of the low-eccentricity $x_1(1)$ and
$x_1^*(2)$ orbits in the full barred potential, which are both $2/1$ periodic
orbits.

\section{$N$-body models}
%%%%%%%%%%%%%%%%%%%%%%%%%
\label{nbody}

In addition to the test particle simulations, we have also run a
high-resolution \mbox{$N$-body} simulation whose predictions can be compared
with the results of the former sections. The main difference of this
simulation with respect to the previous simulations is that it is
three-dimensional and completely self-consistent, i.e. with no rigid component
and allowing the development of spiral arms. The description of the simulation
hereafter is based on physical units in which the initial conditions provide a
reasonable axisymmetric model of the present Milky Way. In these units,
$R_{\hbox{\tiny OLR}}=9$~kpc at $t=2.32$~Gyr.
\par The simulation, also discussed in Fux (\cite{F3}), is similar to the
simulations presented in Fux (\cite{F1}), starting from a bar unstable
axisymmetric model including a nucleus-spheroid, a disc and a dark halo
component with parameters set to $a=1$~kpc,
$M_{\rm NS}=2.6\cdot 10^{10}$~M$_{\odot}$, $h_R=3.2$~kpc, $h_z=300$~pc,
$M_{\rm D}=5.0\cdot 10^{10}$~M$_{\odot}$, $b=9.1$~kpc and
$M_{\rm DH}=2.6\cdot 10^{11}$~M$_{\odot}$ (in the same notation as in Fux
\cite{F1}). It involves $14\,299\,552$ particles of which $5\,553\,784$ belong
to the disc component. The simulation uses the double polar-cylindrical grid
technique described in Fux (\cite{F2}) to solve the gravitational forces, with
$N_R\times N_{\phi}\times N_z=94\times 96\times 253$, $H_z=25$~pc and imposed
reflection symmetry with respect to the plane $z=0$ and the \mbox{$z$-axis}.
The time integrator is a leap-frog with a time step $\Delta t=0.05$~Myr. The
simulation has been carried out until $t=2.65$~Gyr. The phase space
coordinates have been saved every 50~Myr for all particles and every Myr for
the disc particles within a fixed annulus $7.5\leq R/{\rm kpc}\leq 10.5$.
\par After the formation of the bar at about $1.4$~Gyr, the simulation reveals
a complex velocity structure with multiple streams occuring almost everywhere
in the disc. The streams, as well as the velocity dispersions, remain very
time dependent, even within space regions corotating with the bar. Some mpeg
movies showing the time evolution of the $u-v$ distribution for a realistic
position of the Sun are available on the web at the address
{\small \tt http://www.mso.anu.edu.au/\~{}fux/streams.html}.
However, the strong time dependency is certainly partly the consequence of
incomplete phase mixing as in the test particle simulations of the previous
sections. Therefore, we will again average in time the $u-v$ distributions
resulting from the \mbox{$N$-body} simulation to minimise this problem and
thus focus on the average properties of these distributions.
\begin{figure}
\includegraphics[width=8cm]{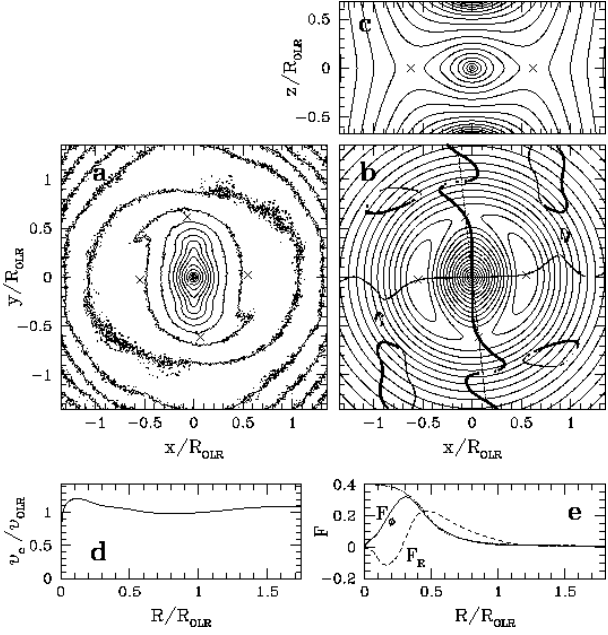}
\caption{\small Some properties of the \mbox{$N$-body} simulation averaged over
$2\leq t/{\rm Gyr}\leq 2.65$ and scaled to a time independent
$R_{\hbox{\tiny OLR}}$ (see text for details). {\bf a)}~Face-on surface
density of the luminous (disc + nucleus-spheroid) particles, with contours
spaced by 0.5 magnitude. {\bf b)}~Effective potential in the plane $z=0$, with
the spacing of the contours increasing by a factor $1.2$ as one moves from
$L_{4/5}$ to lower $\Phi_{\rm eff}$ regions. The curves made of large and
small dots represent respectively the azimuthal minima and maxima of
$\Phi_{\rm eff}$ at local radius. In these first two frames, galactic rotation
is clockwise and the crosses indicate the Lagrangian points $L_{1/2}$ and
$L_{4/5}$. {\bf c)}~Effective potential in the vertical plane passing through
the Lagrangian points $L_1$ and $L_2$ and shown by the dotted straight line in
the former frame. {\bf d)}~Rotation curve of the axisymmetrised average model,
with $v_{\hbox{\tiny OLR}}\equiv v_{\rm c}(R_{\hbox{\tiny OLR}})$.
{\bf e)}~Radial dependence of the azimuthal and radial bar strengths
$F_{\phi}$ and $F_R$. The unlabelled dotted curve is the azimuthal strength of
the analytical barred potential given by Eq.~(\ref{pot}) for $F=0.20$.}
\label{propn}
\end{figure}
\par A complication arises from the fact that the pattern speed of the bar
decreases with time due to the angular momentum the bar loses to the (live)
dark halo, so that the OLR does not lie at a fixed radius anymore but moves
outwards. The decrease of $\Omega_{\rm P}$ is probably not that large in real
disc galaxies with a substantial gas fraction like the Milky Way, as shown by
self-consistent numerical simulations with an SPH component (e.g. Friedli \&
Benz \cite{FB}). Hence, the $u-v$ distributions at a given
$R_{\circ}/R_{\hbox{\tiny OLR}}$ must be computed within space volumes
comoving with $R_{\hbox{\tiny OLR}}$. We found that the evolution of the bar
position angle $\vartheta(t)$ over the time interval
$1.5\leq t/{\rm Gyr}\leq 2.65$ can be well fitted by an analytical function of
the form:
\begin{equation}
\vartheta(t)=\vartheta_{\infty}-\frac{\Omega_{\circ}}{\mu}\exp{(-\mu t)},
\end{equation}
where $\vartheta_{\infty}$, $\Omega_{\circ}$ and $\mu$ are the free
parameters, with residuals in $\vartheta$ never exceeding $5^{\circ}$. From
this we obtain $\Omega_{\rm P}(t)={\rm d}\vartheta/{\rm d}t$ and
$R_{\hbox{\tiny CR}}(t)=v_{\circ}'/\Omega_{\rm P}(t)$, where $v_{\circ}'$ is
matched so that $R_{\hbox{\tiny CR}}=[R(L_{1/2})+R(L_{4/5})]/2$ on the
average. The radius of the OLR is then derived from $R_{\hbox{\tiny CR}}$ via
the flat rotation curve relation, which yields a very good approximation of
$R_{\hbox{\tiny OLR}}$ on the average, and represents the smoothly changing
reference adopted for the distance normalisation. Because the rotation curve
remains nearly flat and constant with time at intermediate $R$, no velocity
scaling is required.
\begin{figure}
\includegraphics[width=8cm]{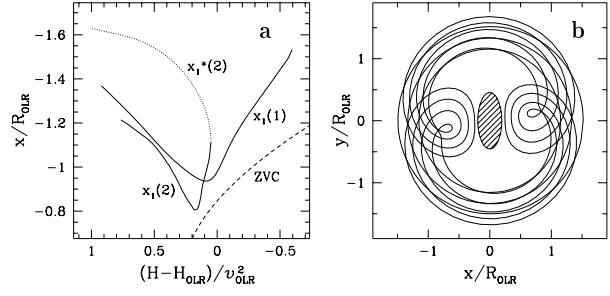}
\caption{\small {\bf a)} Characteristic diagram for the low-eccentricity $x_1(1)$,
$x_1^*(2)$ and $x_1(2)$ periodic orbits in the time averaged \mbox{$N$-body}
model. The full and dotted lines indicate stable and unstable orbits
respectively, and the dashed line the zero velocity curve.
$v_{\hbox{\tiny OLR}}$ is the circular velocity at the OLR in the
axisymmetrised potential. All curves are interrupted at high~$H$ (see text),
and the $x_1(1)$ curve is truncated on the low-$H$ side at the outer $1/1$
resonance. {\bf b)} Sequence of $x_1(1)$ orbits in the $x-y$ plane for the
same model. The orbits are sampled at constant Hamiltonian interval and cover
the whole range of the corresponding characteristic curve in the former
diagram. The orientation of the bar is sketched by the shaded ellipse and its
rotation is clockwise.}
\label{orbchar}
\end{figure}
\par The time average is done over the interval $2\leq t/{\rm Gyr}\leq 2.65$
and as underlying mass distribution and potential associated with the average
velocity distributions, we take those resulting from the sum of the $50$~Myr
spaced output models of the simulation within the same time interval and with
the mass in each model rescaled as the distances to preserve the velocity
scale. The number of models added together is rather low (only 14), yielding
only a crude estimate of the true average quantities, especially regarding
azimuthal variations in the spiral arm regions. Figure~\ref{propn} shows some
properties of the resulting average model. At given radius $R$, the azimuthal
and radial bar strengths $F_{\phi}(R)$ and $F_R(R)$ (Fig.~\ref{propn}e) are
defined as the most extreme values over azimuth of
$|A_{\phi}(R,\phi)/A_R^{\rm axisym}(R)|$ and
$[A_R(R,\phi)-A_R^{\rm axisym}(R)]/A_R^{\rm axisym}(R)$ respectively, where
$A_{\phi}$ and $A_R$ are the azimuthal and radial accelerations in the plane
$z=0$ and $A_R^{\rm axisym}$ is the axisymmetric part of $A_R$. In particular,
$F_{\phi}(a=R_{\hbox{\tiny CR}}/1.25)$ coincides with the definition of bar
strength used in the former sections. The time averaged model has
$F_{\phi}(a)\approx 0.2$, i.e. a rather strong bar. However, Buta \&
Block (\cite{BB}) have introduced a bar strength classification in terms of a
parameter $Q_{\rm b}$, corresponding here to the radial maximum of
$F_{\phi}(R)$, and present galaxies with values of this parameter up to
$0.65$. Our average model has $Q_{\rm b}\approx 0.325$ and thus falls only in
the middle of the range covered by real galaxies.
\begin{figure*}
\vspace{12.2cm}
\caption{\small Time averaged planar velocity distribution of the disc particles in
the $N$-body simulation as a function of position in space and within
vertical cylinders of radius $R_{\circ}/80$. The horizontal and vertical axes
of the frames are $v/v_{\circ}$ and $u/v_{\circ}$ respectively, with
$v_{\circ}$ being the circular velocity at local radius in the corresponding
axisymmetriesed average model. The velocity contours are as in
Fig.~\ref{pop10}, as well as the $H$-contours (drawn at $z=0$), the resonance
curves and the traces of the periodic orbits, which all refer to the average
rotating potential. Only orbits from the $x_1(1)$, $x_1^*(2)$ and $x_1(2)$
families and within the Hamiltonian range scanned by the characteristic curves
in Fig.~\ref{orbchar}a are indicated.}
\label{ncorps}
\end{figure*}
\par The disc scale length between corotation and slightly outside the OLR
(Fig.~\ref{propn}a) has substantially increased with respect to the initial
conditions, in agreement with the test particle results. The effective
potential (Fig.~\ref{propn}b) indicates Lagrangian points which significantly
lag the bar principal axes. This is the effect of the spiral arms starting at
the end of the bar, which were absent in the test particle simulations and now
produce a twist of the potential well. Note however that, as pointed out by
Zhang (\cite{ZX}), there is a phase shift between the potential and the
density wells, with the former leading the latter outside the bar. The
characteristic curves of the main periodic orbit families and the $x-y$
configuration of the $x_1(1)$ orbits in the average potential are given in
Fig.~\ref{orbchar}. The characteristic curves are truncated at large $H$,
where they start to bend and interfere with each other presumably as a
consequence of the sharp phase change in the potential well at large radii
(see Fig.~\ref{propn}b). The $x_1(1)$ orbits respond to the twisted potential
well by having their major axis departing more and more from the
\mbox{$y$-axis} as $H$ decreases, and the closed orbits of the other families
share a similar response. Since the cusps of the cusped orbits now occur away
from the coordinate axes, the characteristic curves no longer reach the ZVC.
\par The time averaged $u-v$ distributions are presented in Fig.~\ref{ncorps}.
The diagrams are computed by summing the $1$~Myr spaced outputs of the
simulation, yielding much more accurate time averages than for the properties
based on the $50$~Myr spaced outputs, and using the same space volumes,
velocity binning and smoothing procedure as for the test particle simulations.
However, contrary to the latter simulations, the diagrams are based on a
unique simulation and therefore the initial conditions for each diagram now
scale as $R_{\hbox{\tiny OLR}}$ instead of $R_{\circ}$. Hence the pattern
speed and size of the bar are not the only parameters that change for
different values of $R_{\circ}/R_{\hbox{\tiny OLR}}$. In particular, the
initial velocity dispersions decrease with $R$, causing a similar trend in the
final velocity distributions. Diagrams have been derived at an azimuthal
interval $\Delta \varphi =10^{\circ}$, but only those at $30^{\circ}$ spacing
are shown.
\par A priori, some properties inferred from the test particle simulations
seem less robust in the \mbox{$N$-body} simulation: the non-resonant $x_1(1)$
orbits are somewhat less coincident with peaks in the velocity distributions,
and the resonant $x_1(1)$ orbits, i.e. those with traces on the $2/1$
resonance curve, are less correlated with depleted $u-v$ regions at high
eccentricities. Moreover, the low angular momentum peak often lies well inside
the hot orbit region even when it is still mostly associated with regular
\mbox{quasi-$x_1(2)$} orbits, as indicated by the presence of a stable
low-eccentricity $x_1(2)$ orbit. In fact, a closer inspection reveals that the
velocity distributions in the \mbox{$N$-body} simulation at given
$R_{\circ}/R_{\hbox{\tiny OLR}}$ appear to best match those of the test
particle simulations at a typically $10$\% larger value of
$R_{\circ}/R_{\hbox{\tiny OLR}}$. This can be explained by a delayed response
of the phase space density distribution to the growing absolute OLR radius:
the $u-v$ distributions do not instantaneously adjust to the moving
$R_{\hbox{\tiny OLR}}$ and therefore always reflect an earlier smaller
$R_{\hbox{\tiny OLR}}$ value instead of the current one. Hence the velocity
distributions in Fig.~\ref{ncorps} should virtually be shifted upwards by
roughly one frame to be more consistent with the
$R_{\circ}/R_{\hbox{\tiny OLR}}$ scale and the other plotted informations.
Actually, the effective $R_{\hbox{\tiny OLR}}$ must be a function of the
orbits.
\par With such a correction in mind, the \mbox{$N$-body} velocity
distributions now share much more the same properties as in the test particle
simulations. The low angular momentum mode, when purely induced by chaotic
orbits as in the frame $R_{\circ}/R_{\hbox{\tiny OLR}}=1.1$ and
$\varphi =30^{\circ}$ of Fig.~\ref{ncorps}, also no longer peaks outside the
$H_{45}$ contour, but rather between the $H_{12}$ and the $H_{45}$ contours.
The velocity distributions most resemble those of the test particle
simulations with a bar strength $F=0.15$ (not shown in this paper, but see
Fux \cite{F4}). Since $F\approx 0.20$ in our average \mbox{$N$-body} model,
this suggests that the velocity distributions are less responsive to the bar
strength in the more realistic 3D \mbox{$N$-body} simulation than in the 2D
test particle simulations.
\par The symmetries found in Sect.~\ref{testpart} for the velocity
distributions, and in particular the $u$-symmetry at $\varphi=0$ and
$\varphi=90^{\circ}$, obviously break and the velocity distributions in the
\mbox{$N$-body} model at given $\varphi$ and fixed effective radius seem to
compare best with the corresponding distributions in the test particle
simulations at an angle $\sim \varphi-10^{\circ}$, suggesting that the velocity
distributions know about the spiral arm induced average local twist of the
potential well relative to the bar major axis. While this is especially true
for the more regular low-$H$ regions, the velocity distributions show no
significant phase shift at all in the hot orbit region. The reason is because
the hot orbits are more eccentric and therefore are sensitive to more inner
features of the potential. It should be noted that in \mbox{$N$-body}
simulations like the one presented here, spiral arms are particularly strong
during about $1$~Gyr after the formation of the bar, so that the reported
effects of spiral arms are probably overestimates for the Milky Way if the
Galactic bar is old.
\begin{figure}
\includegraphics[width=7cm]{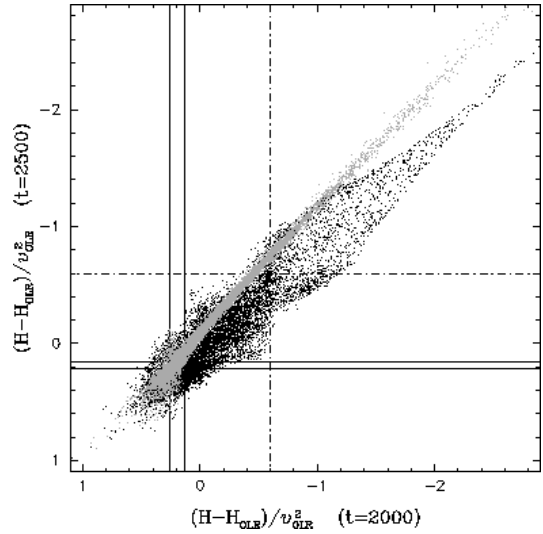}
\caption{\small Comparison of the normalised Hamiltonian values of a random selection
of $2\cdot 10^5$ disc component particles in the \mbox{$N$-body} simulation at
times $t=2000$ and $2500$~Myr. $v_{\hbox{\tiny OLR}}$ and
$H_{\hbox{\tiny OLR}}$ respectively are the velocity and Hamiltonian value of
the circular orbit at the OLR of the axisymmetrised potential. The points in
grey represent the particles inside the circle passing through the Lagrangian
points $L_{1/2}$ at $t=2000$~Myr. These particles exist down to and beyond the
smallest displayed value of $H$, as the effective potential at the central
Lagrangian point is
$(\Phi_{\rm eff}(L_3)-H_{\hbox{\tiny OLR}})/v_{\hbox{\tiny OLR}}\approx -5.2$.
The solid lines indicate the normalised $H_{12}$ and $H_{45}$ values and the
dash-dotted lines the normalised $H$-value of the circular orbit at the outer
$1/1$ resonance (in the axisymmetrised potential).}
\label{consham}
\end{figure}
\par A potentially important difference of 3D models with respect to 2D models
is that the effective potential a star experiences near corotation depends on
its distance from the Galactic plane (see Fig.~\ref{propn}c). This raises the
average value of the Jacobi integral required for the stars to cross the
corotation radius and thus renders such a crossing more difficult. For stars
on the Solar circle, the higher effective value of $H_{12}$ is not compensated
by their departure from the Galactic plane or a $w$ velocity component.
Indeed, in our average \mbox{$N$-body} model, the change of effective
potential near corotation when moving from $z=0$ to $z=300$~pc is
$\Delta \Phi_{\rm eff}/v_{\hbox{\tiny OLR}}^2\approx 0.233$ (with
$v_{\hbox{\tiny OLR}}$ as defined in the caption of Fig.~\ref{orbchar}), while
this change at the OLR of the axisymmetrised potential is only $0.004$, and
adding a vertical velocity of $w/v_{\hbox{\tiny OLR}}=0.1$ only increases
$H/v_{\hbox{\tiny OLR}}^2$ by $0.005$. Hence 2D models certainly exaggerate
the traffic of stars on hot orbits travelling from one side of corotation to
the other.
\par Finally, Fig.~\ref{consham} shows how the value of the Hamiltonian is
conserved during the \mbox{$N$-body} simulation. The main result is that the
$H$-values at different times are much better related for bar particles than
for (the~dynamically defined) disc and hot particles. In particular, bar
particles remain bar particles, but disc particles can easily transform into
hot particles and vice versa for
$(H-H_{\hbox{\tiny OLR}})/v_{\hbox{\tiny OLR}}^2\ga -0.2$, supporting the
presumption in Sect.~\ref{testpart} that spiral arms may induce exchanges
between regular and chaotic phase space regions in real galaxies. Near the
$1/1$ resonance, the disc particles also reveal a smaller scatter of their
past versus present $H$ relation. The normalised Hamiltonian (as well as the
absolute non-rescaled Hamiltonian) increases on the average for the disc
particles, which may be understood by the fact that the contribution of the
term $-\Omega_{\rm P}^2R^2/2$ to $\Phi_{\rm eff}$ diminishes as the bar
rotates slower, and decreases for the bar particles, owing to the deepening
of the central potential well.
%deltaphi(Rolr,300pc,axisym)=1.63853E-04, deltaphi(R12,300pc,bar)=1.02939E-02
\par It would be interesting to investigate the evolution of the particle
$H$-values in an \mbox{$N$-body} simulation with a bar rotating at a constant
frequency, for example without including a live dark halo component, in order
to disentangle the effects of the spiral arms from the effects of a decreasing
pattern speed. It would be also worth to explore the diffusion of particles
from regular to chaotic phase space regions and vice versa, starting with 2D
\mbox{$N$-body} simulations in a first approach. One problem to be clarified
is why the \mbox{$N$-body} velocity distributions look so similar to the test
particle ones, despite the action of such a diffusion process. A detailed
analysis of the orbital structure in 3D models remains to be done, and in
particular of the properties of the vertical motion within regular and chaotic
regions.

\section{Models versus observations}
%%%%%%%%%%%%%%%%%%%%%%%%%%%%%%%%%%%%
\label{compar}

Before concluding, we now present a selection of test particle and
\mbox{$N$-body} velocity distributions yielding a good match to the observed
one, confront the \mbox{quasi-$x_1(2)$} orbit and chaotic orbit
interpretations of the Hercules stream, paying also attention to the case of
the Hyades stream, and discuss how the models could be further improved.
\par Beside the parameters in the initial conditions of the simulations, the
free model parameters are $R_{\circ}/R_{\hbox{\tiny OLR}}$, $\varphi$, the
velocity scale specified by $v_{\circ}$ (defined as the local circular
velocity in the axisymmetric part $\Phi_{\circ}$ of the potential for the
\mbox{$N$-body} simulation), which should lie between $180$ and
$230$~km\,s$^{-1}$ (e.g. Sackett \cite{PS}), and the velocity of the Sun
$(v_{\rm s},u_{\rm s})$ relative to the circular orbit in $\Phi_{\circ}$.
A~commonly adopted velocity reference in the Solar neighbourhood is the LSR,
defined as the velocity of the most nearly circular closed orbit passing
through the present location of the Sun according to Binney \& Merrifield
(\cite{BM}). This definition, which is merely an attempt to generalise the
circular LSR orbit of the axisymmetric case to non-axisymmetric potentials, is
not always well adapted. The most reasonable LSR orbit candidates near the OLR
of a barred potential indeed are the stable low-eccentricity $x_1(1)$ and
$x_1(2)$ orbits, but some space positions near the OLR circle are visited by
neither of these orbits in our models (see for example
$R_{\circ}/R_{\hbox{\tiny OLR}}=1.0$ and $\varphi =30^{\circ}$ in
Fig.~\ref{pop10}). However, for $R_{\circ}/R_{\hbox{\tiny OLR}}\ga 1.0$, there
always exists a prominent peak of low-eccentricity \mbox{quasi-$x_1(1)$}
orbits in the model velocity distributions, which, as pointed out in
Sect.~\ref{testpart}, not necessarily coincides with the trace of the
non-resonant $x_1(1)$ orbit when there is one. The maximum of this peak will
therefore be taken as the model ``LSR'' and will be preferably associated to
the Coma Berenices stream, which is the local maximum in the observed velocity
distribution that lies closest to the heliocentric velocity of the LSR
$(v,u)_{\rm LSR}\approx (-5,10)$~km\,s$^{-1}$ derived from the Hipparcos data
(Dehnen \& Binney \cite{DB}).
\par For $R_{\circ}\ga R_{\hbox{\tiny OLR}}$, the \mbox{quasi-$x_1(1)$} peak
is always close to the circular orbit of the axisymmetrised potential, except
near the OLR radius and $\varphi =90^{\circ}$ where it reaches a maximum
positive $v$-offset of $\sim 0.05v_{\circ}$ for all explored bar strengths.
Under the above circumstances and for realistic space positions, the azimuthal
velocity of the Sun should exceed the circular velocity by
$5-10$~km\,s$^{-1}$.
\par
\begin{figure*}
\includegraphics[width=17.5cm]{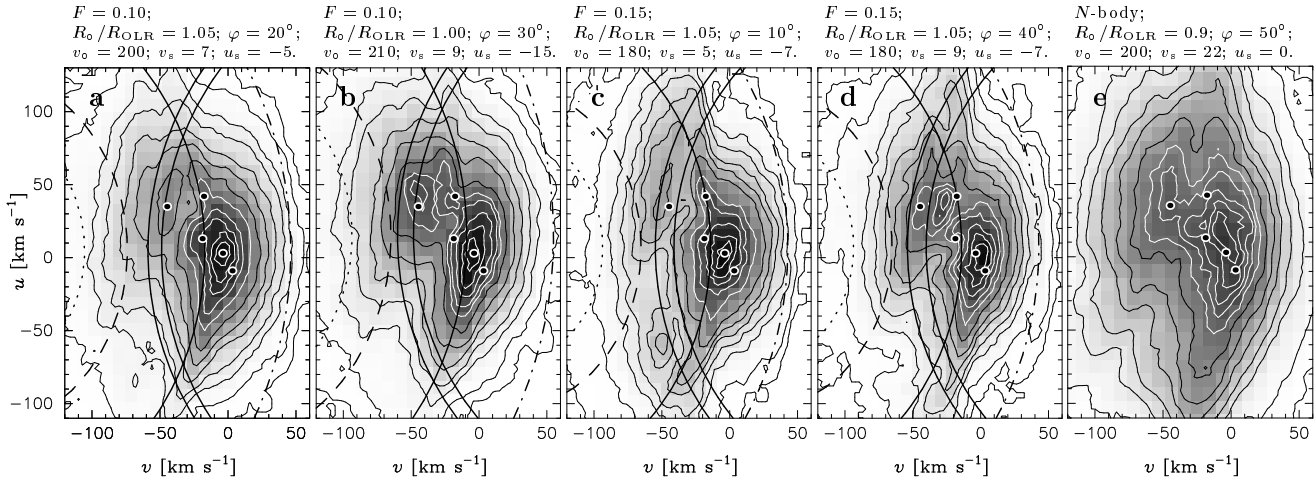}
\caption{\small Selection of scaled velocity distributions from the test particle
(frame~{\bf a} to {\bf d}) and \mbox{$N$-body} (frame~{\bf e}) simulations,
with the various parameters indicated on the top of the frames and the
velocity origin at the adopted Solar motion. The velocity window and the
velocity contours are the same as in Fig.~\ref{obs}. The filled circles
represent the mean velocities of the streams listed in Table~\ref{stream},
excluding the Arcturus stream. All distributions from the test particle
simulations are time averages over $55\leq t/t_{\rm b}\leq 65$. The
$H$-contours and the resonance curves are as in Fig.~\ref{lia10} and are not
plotted for the \mbox{$N$-body} model because of the time delay problem
discussed in Sect.~\ref{nbody}.}
\label{best}
\end{figure*}

 The selected model velocity distributions are displayed in
Fig.~\ref{best}. The distributions are derived according to the same
procedures as described in Sects.~\ref{testpart} and~\ref{nbody}. Frame~(a)
shows a case where the Hercules-like stream is induced exclusively by chaotic
orbits and peaks {\it inside} the $H_{12}$ contour, illustrating the fact that
chaotic modes not necessarily only occur in the hot orbit region. Here the
Hyades stream coincides with a chaotic overdensity associated with a narrow
and low-$H$ chaotic breach roughly along the OLR curve, i.e. an interpretation
similar to the one proposed in Sect.~\ref{testpart} for the $u<0$ extension of
the LSR mode. Frame~(c) gives a case where the Hercules-like stream now falls
entirely in the hot orbit region and where the Hyades stream has the same
chaotic origin as in frame~(a).
\par Frame~(e), derived from the \mbox{$N$-body} simulation and presenting a
larger velocity dispersion, provides a remarkable example of a Hercules-like
stream sustained exclusively by \mbox{quasi-$x_1(2)$} orbits. The test
particle simulations develop \mbox{quasi-$x_1(2)$} modes which cannot be as
easily matched to the Hercules stream in our scaling procedure, generally
falling right between this stream and the Hyades stream. This can be explained
by the different local slope of the circular velocity $v_{\rm c}$ in the
\mbox{$N$-body} and the test particle models. As explained by D2000 in terms
of orbital frequencies, the separation between the \mbox{quasi-$x_1(1)$} and
the \mbox{quasi-$x_1(2)$} modes increases with
${\rm d}v_{\rm c}/{\rm d}R(R_{\circ})$.
Since the average \mbox{$N$-body} model has a slightly raising rotation curve
near the OLR radius (Fig.~\ref{propn}d), its circular velocity gradient is
larger than for the flat rotation curve test particle simulations and thus the
\mbox{quasi-$x_1(2)$} mode is found at higher asymmetric drift relative to the
\mbox{quasi-$x_1(1)$} mode. However, the fact that observations support a
rather gently declining rotation curve at $R_{\circ}$ and that a large
inclination angle of the bar is needed ($\varphi\ga 40^{\circ}$ for
$R_{\circ}>R_{\hbox{\tiny OLR}}$) are arguments against the
\mbox{quasi-$x_1(2)$} interpretation of the Hercules stream. On the other
hand, the displacement of the \mbox{quasi-$x_1(2)$} peak towards the $H_{12}$
contour with increasing integration time mentioned in Sect.~\ref{testpart} for
$F=0.10$ and $R_{\circ}\approx R_{\hbox{\tiny OLR}}$ may reinforce this
interpretation.
\par Frame~(d) is an example with two distinct low angular momentum peaks, the
most negative $v$ one being related to chaotic hot orbits and the other one to
\mbox{quasi-$x_1(2)$} orbits. It would be interesting to check whether a
sufficiently negative ${\rm d}v_{\rm c}/{\rm d}R(R_{\circ})$ is able to shift
the \mbox{quasi-$x_1(2)$} mode more towards the true location of the Hyades
stream and thus yield a model velocity distribution with a better overall
match to the observed one. Note that the chaotic orbit mode will not
necessarily be shifted as the \mbox{quasi-$x_1(2)$} mode, because its location
in the $u-v$ plane does not actually depend on the local slope of the circular
velocity, but rather on the difference of $\Phi_{\rm eff}$ between the current
space position and the Lagrangian point $L_{1/2}$, which determines the $u-v$
location of the $H_{12}$ contour\footnote{In particular, at
$R_{\hbox{\tiny OLR}}\approx 1.1$ and $\varphi =25^{\circ}$, where the low
angular momentum mode has a chaotic origin, the $v$ squashing of the
bimodality reported by D2000 when decreasing his rotation curve parameter
$\beta$ is perhaps not the consequence of a {\it local} change of the circular
velocity slope, but of an implied lower relative value of
$\Phi_{\rm eff}(L_{1/2})$.}. Finally, frame~(b) displays a surprising case
where the velocity distribution in the \mbox{quasi-$x_1(2)$} region of the
$u-v$ plane (see Fig.~\ref{uvx1}) seems to have split into two peaks
coinciding well with the locations of the Hercules and Hyades streams, i.e.
both these streams have a \mbox{quasi-$x_1(2)$} origin.~However, this is
likely to be a transitory situation resulting from the unachieved phase mixing
near $R_{\circ}=R_{\hbox{\tiny OLR}}$ (see Sect.~\ref{testpart}).
\par These examples illustrate the variety of possible interpretations for the
Hercules and Hyades streams, and it is very hard at this stage to decide with
certainty which one is the most appropriate. The splitting of the LSR mode
into the Pleiades, Coma Berenices, Sirius and other streams is probably not
related to the bar itself and has a more local origin, like for instance the
effect of time dependent spiral arms.

\section{Conclusion}
%%%%%%%%%%%%%%%%%%%%
\label{concl}

The Galactic bar induces a characteristic splitting of the disc phase space
into regular and chaotic orbit regions, with the latter regions owing only to
the non-axisymmetric part of the potential in the limit of no vertical motion.
In this paper, we have isolated these two kind of regions, as well as the
quasi-periodic orbit sub-regions inside the regular regions associated with
the stable $x_1(1)$ and $x_1(2)$ periodic orbits respectively, within the same
analytical 2D rotating barred potential with flat azimuthally averaged
rotation curve as in D2000. We then have run test particle simulations in this
potential and a more realistic self-consistent 3D \mbox{$N$-body} simulation
to find out how the disc distribution function outside the bar region relates
to such a phase space splitting and in particular how chaos may explain
features in the Solar neighbourhood stellar kinematics like the Hercules
stream.
\par Beside the bar strength, the regular versus chaotic splitting of phase
space, investigated via the largest Liapunov exponent, is mainly determined by
the value of the Hamiltonian~$H$ (or Jacobi's integral) and by the bar related
resonances. In two dimensions and at fixed space position, the constant-$H$
contours in the galactocentric $u-v$ velocity plane are circles centred on
$(v,u)=(R_{\circ}\Omega_{\rm P},0)$ and of radius
$\sqrt{2(H-\Phi_{\rm eff}^{\circ})}$, where $R_{\circ}$ is the galactocentric
distance, $\Omega_{\rm P}$ the rotation frequency of the bar and
$\Phi_{\rm eff}^{\circ}$ the local effective potential. The fraction of
chaotic orbits increases with $H$ and there is a sharp average transition from
regular to chaotic behaviour in the $u-v$ plane when crossing the contour
corresponding to the effective potential at the saddle Lagrangian points,
$H_{12}\equiv \Phi_{\rm eff}(L_{1/2})$, which generally intersects the
\mbox{$v$-axis} at lower velocity than the circular orbit in the axisymmetric
part of the potential. At $H<H_{12}$, the orbits are rather regular, while at
$H>H_{12}$, which defines the category of {\it hot orbits} susceptible to
cross the corotation radius, they are rather chaotic.
\par The resonances, on the other hand, generate an alternation of regular and
chaotic orbit arcs in the velocity plane which, contrary to the low-$v$ part
of the \mbox{$H$-contours}, are opened towards lower angular momentum. At bar
inclination angles $\varphi=0$ and $\varphi=90^{\circ}$, the maxima or minima
of these stochasticity waves are in phase with the resonance curves derived
from the axisymmetric limit and the arcs are symmetric in $u$, reflecting the
four-fold symmetry of the potential. At intermediate angles, these extrema
become offset with respect to the resonance curves and the $u$-symmetry
breaks. In particular, at $R\ga R_{\hbox{\tiny OLR}}$ and
$\varphi \sim 30^{\circ}$, a prominent regular region of eccentric
\mbox{quasi-$x_1(1)$} orbits extends well within the hot orbit region at
$u\la 0$, while the $u>0$ counterpart of the OLR curve is surrounded by a wide
chaotic region consistent with the location of the Hercules stream.
\par For a moderate bar strength ($F=0.10$), the low-eccentricity and
non-resonant \mbox{quasi-$x_1(2)$} orbit regions exist only for
$R/R_{\hbox{\tiny OLR}}\la 1.1$ and over an angle range around
$\varphi=90^{\circ}$ increasing towards smaller $R$. There is no such region
near the default position considered in D2000, i.e.
$R/R_{\hbox{\tiny OLR}}\approx 1.1$ and $\varphi =25^{\circ}$, compromising
the \mbox{quasi-$x_1(2)$} orbit interpretation given by Dehnen for the
Hercules-like mode occuring in his simulations at the most realistic positions
of the Sun relative to the bar.
\par The test particle simulations, started from axisymmetric initial
conditions and progressively exposed to the full rotating barred potential,
reveal a decoupled evolution of the disc distribution function within the
regular and chaotic phase space regions. In the regular regions, the phase
space density after phase mixing is roughly the same as the initial one,
whereas in the chaotic regions, the particles quickly evolve towards a uniform
population of the easily available phase space volume via chaotic mixing,
resulting in a substantial density re-adjustment. Because the space region
within corotation represents a large initial reservoir of hot chaotic orbit
particles which are spread throughout the disc by this process, yielding a net
outward migration of such particles, the chaotic regions in the $u-v$ plane
outside corotation become more heavily crowded than the regular regions at
$H\ga H_{12}$. In particular, the wide and predominantly $u>0$ chaotic region
mentioned above for realistic space positions of the Sun appears as an
overdensity in the $u-v$ distribution, providing a coherent interpretation of
the Hercules stream and explaining the $u>0$ property of this stream.
According to this interpretation, the Hercules stream involves stars on
essentially chaotic orbits which are forced to avoid the regular $x_1(1)$
region at negative $u$.
\par The time averaged disc $u-v$ velocity distributions inferred from the
\mbox{$N$-body} simulation are remarkably similar to those of the test
particle simulations, despite the action of the transient spiral arms which
allows at least some of the particles to diffuse from the regular to the
chaotic regions and vice versa. At low eccentricity, the orbits are less
sensitive to the inner features of the potential and the azimuthal properties
of the velocity distributions essentially align with the average local phase
shift of the potential well relative to the bar major axis induced by the
spiral arms.
\par The velocity distributions may be very time dependent if for
instance the bar has formed recently, because of the phase mixing occuring
in the disc during at least $\sim 10$ bar rotations after the growth of the
bar according to the test particle simulations. The $u-v$ distributions in the
\mbox{$N$-body} simulation at fixed space position relative to the bar also
display a strong temporal behaviour (see the mpeg movies at
{\small \tt http://www.mso.anu.edu.au/\~{}fux/streams.html}), as expected from the
presence of the transient spiral waves. However, since the simulation has been
run for only about $1.25$~Gyr after the formation of the bar, phase mixing
is still operating in the disc component, rendering difficult to disentangle
from it the individual effects of such waves. The $N$-body simulation also
gives some insight on the consequences of evolving bar parameters: the slowly
decreasing pattern speed of the bar mainly introduces a delayed response of
the disc distribution function to the outward moving resonances, so that the
velocity distributions at a given time rather reflect a higher value of
$R/R_{\hbox{\tiny OLR}}$ than the true instantaneous one when compared with
the constant $\Omega_{\rm P}$ test particle simulations. For completeness, one
should mention that other perturbations than the bar and the spiral arms may
provide alternative explanations of the local stellar streams, like for
example the interactions of the Milky Way with its satellite companions.
\par Finally, the process of chaotic mixing, combined with the possible
stellar exchanges between the regular and chaotic phase space regions
resulting from the diffusion of stars by transient spiral arms or molecular
clouds, may provide an new and efficient way of heating galactic discs which
remains to be explored.

%%%%%%%%%%%%%%%%%%%%%%%%%%%%%%%%%%%%%%%%%%%%%%%%%%%%%%%%%%%%%%%%%%%%%%
\begin{acknowledgements}
I would like to thank K.C.~Freeman for a careful reading of the manuscript and
A.~Kalnajs for many enriching interactions. I am also thankful to
Walter Dehnen for having partly inspired the present investigation and for
several enlightening discussions, and to the University of Geneva where the
\mbox{$N$-body} simulation presented in Sect.~\ref{nbody} has been performed.
This work was mainly supported by the Swiss National Science Foundation.
\end{acknowledgements}
%%%%%%%%%%%%%%%%%%%%%%%%%%%%%%%%%%%%%%%%%%%%%%%%%%%%%%%%%%%%%%%%%%%%%%

\end{document}